\documentstyle[11pt,aaspp4]{article}
\def\gax{\mathrel{\raise.3ex\hbox{$>$}\mkern-14mu\lower0.6ex\hbox{$\sim$}}}
\def\lax{\mathrel{\raise.3ex\hbox{$<$}\mkern-14mu\lower0.6ex\hbox{$\sim$}}}
\def\gtorder{\mathrel{\raise.3ex\hbox{$>$}\mkern-14mu
             \lower0.6ex\hbox{$\sim$}}}
\def\ltorder{\mathrel{\raise.3ex\hbox{$<$}\mkern-14mu
             \lower0.6ex\hbox{$\sim$}}}

\input psfig.sty

\begin{document}

\title{Tests for Substructure in Gravitational Lenses}

\author{C.S. Kochanek$^{(a)}$ and N. Dalal$^{(b)}$\altaffilmark{1}}
\affil{$^{(a)}$ Harvard-Smithsonian Center for Astrophysics, 60 Garden
        St., Cambridge, MA 02138}
\affil{email: ckochanek@cfa.harvard.edu}
\affil{$^{(b)}$ School of Natural Sciences, Institute for Advanced Study,
	Einstein Drive, Princeton NJ 08540}
\affil{email: neal@ias.edu}
\altaffiltext{1}{Hubble Fellow}

\begin{abstract}
The flux anomalies in four-image gravitational lenses can be interpreted
as evidence for the dark matter substructure predicted by cold dark matter 
(CDM) halo models. In principle, these flux anomalies could arise from alternate 
sources such as absorption, scattering or scintillation by the interstellar
medium (ISM) of the lens galaxy, problems in the ellipsoidal macro models used 
to fit lens systems, or stellar microlensing.  
We apply several tests to the data that appear to rule out these alternate
explanations.  First, the radio flux anomalies show no significant dependence 
on wavelength, as would be expected for almost any propagation effect in 
the ISM or microlensing by the stars.  
Second, the flux anomaly distributions show the characteristic demagnifications 
of the brightest saddle point relative to the other images 
expected for low optical depth substructure,
which cannot be mimicked by either the ISM or problems in the macro models.
Microlensing by stars also cannot reproduce the suppression of the bright saddle
points if the radio source sizes are consistent with the Compton limit for their
angular sizes.
Third, while it is possible to change the smooth lens models to fit the
flux anomalies in some systems, we can rule out the necessary changes in
all systems where we have additional lens constraints to check the models.
Moreover, the parameters of these models are inconsistent with our present
observations and expectations for the structure of galaxies.  We conclude
that low-mass halos remain the best explanation of the phenomenon.  
\end{abstract}


\section{Introduction}

In Dalal \& Kochanek (\cite{dk02}, DK02 hereafter) we demonstrated that the incidence
of anomalous flux ratios in gravitational lenses was consistent with the 
expected mass fraction of satellites found in cold dark matter (CDM) 
halo simulations (e.g.
Kauffmann~\cite{Kauffmann93}, Moore et al.~\cite{Moore99}, 
Klypin et al.~\cite{Klypin99}, Bode, Ostriker \& Turok~\cite{Bode01},
Zentner \& Bullock~\cite{Zentner02}).  Our quantitative
estimate for the substructure mass fraction was part of
an outburst of interest in substructure and anomalous flux ratios
of specific lenses (e.g. Mao \& Schneider~\cite{Mao98},
Keeton~\cite{Keeton01}, Bradac et al.~\cite{Bradac02},
Chiba~\cite{Chiba02}, Metcalf~\cite{Metcalf02}), 
in CDM models (e.g. Metcalf \& Madau~\cite{Metcalf01},
Chiba~\cite{Chiba02}, Zentner \& Bullock~\cite{Zentner02}),  
and the difficulty in explaining them based
on the properties of the primary lens galaxy (e.g. Metcalf \& Zhao~\cite{Metcalf01},
Keeton, Gaudi \& Petters~\cite{Keetonetal02}, Evans \& Witt~\cite{Evans02},
M\"oeller, Hewett \& Blain~\cite{Moeller02},
Quadri, M\"oeller \& Natarajan~\cite{Quadri02}).  
In DK02
we argued that alternate explanations to substructure for the anomalous
flux ratios were unlikely, but we did not explore the alternate possibilities
in detail or develop tests to distinguish substructure from the alternatives.
We do so in the current paper.  

The alternatives to substructure fall into three broad categories: propagation
effects in the interstellar medium (ISM) of the lens galaxy, problems in the
``macro'' models for the gravitational potential of the lens galaxy, and
confusing ``microlensing'' by the stars in the lens galaxy with the effects
of more massive satellites.  
In this paper we will consider all three of these possibilities. In
\S2 we study the wavelength dependence of the flux anomalies and
develop a simple statistical test to demonstrate that the anomalous
flux ratio problem must be due to gravity.  In \S3 we show that
deviations from elliptical shapes appear unable to account for flux
anomalies, and that errors in the model for the potential of the
primary lens have statistical properties differing from those
created by substructure and observed in the data.
In \S4 we show that it is difficult to explain the anomalous flux ratios of 
the radio lenses using microlensing.  We summarize our results and
outline further tests in \S5.

\section{Ruling out the ISM}

The ISM can affect flux ratios either through absorption or scattering.  In
optical/IR observations, many gravitational lenses show wavelength-dependent
flux ratios consistent with the effects of dust extinction (e.g. Falco et al. 
\cite{Falco99}).  In the radio, some lenses show the effects of scatter broadening
by electrons at low frequencies (e.g. B1933+503, Marlow et al.~\cite{Marlow99a}), 
although in none of these systems is there evidence for
a net change in the radio flux.  
In this section we examine whether the radio lenses show any 
frequency dependent changes in their flux ratios that could be
a signature for the ISM modifying the observed fluxes, and develop
a simple, non-parametric test to distinguish the effects of the ISM
from the effects of gravity. 

\subsection{The Frequency Dependence of Anomalous Flux Ratios \label{sec:opdepth}}

Almost all mechanisms by which the ISM can modify flux ratios should
show frequency-dependent effects, since most scattering
processes relevant to the fluxes of radio images become weaker
at higher frequencies.  For example, weak scintillation causes flux
perturbations with a scaling (assuming Kolmogorov turbulence)
$\propto \nu^{-17/12}$ (Narayan~\cite{narayan92}) and the optical depth
for free-free absorption scales roughly as $\propto \nu^{-2.1}$
(e.g. Mezger \& Henderson~\cite{Mezger67}).  Thus, one  
way to constrain the ISM as an explanation of the anomalous
flux ratios is to examine the allowed frequency dependence of 
the effect.  If the source has
an intrinsic spectrum $f_{\nu,s}$ as a function of frequency
$\nu$, then in the absence of any perturbations from the ISM, 
the images have spectrum $f_{\nu,i}=|M_i| f_{\nu,s}$ given the
(signed) image magnification $M_i$.  If the ISM modifies the
fluxes through a frequency-dependent optical depth $\tau_{\nu,i}$,
then the image fluxes become $f_{\nu,i}=|M_i| f_{\nu,s}\exp(-\tau_{\nu,i})$.

We examined the optical depth function using the four (or more) image radio lenses 
with fluxes measured at 5~GHz and either 8~GHz or 15~GHz 
(MG0414+0534, Katz, Moore \& Hewitt~\cite{Katz97}; 
B0712+472, Jackson et al.~\cite{Jackson98};
B1359+154, Myers et al.~\cite{Myers99};
B1422+231, Patnaik \& Narasimha~\cite{Patnaik01}; 
B1555+375, Marlow et al.~\cite{Marlow99b};
B1608+656, Fassnacht {\it private communication};
B1933+503, Sykes et al.~\cite{Sykes98};
and B2045+265, Fassnacht et al.~\cite{Fassnacht99}).
We used the same models and magnification estimates as were used to
estimate the substructure fraction with a 5\% lower bound on the
flux errors.  The simplest way to illustrate 
the problem is to assume
that one image is strongly affected by the ISM while the others 
suffer from little or no absorption or scattering.  For each image
in a lens, we used the fluxes of the other three to estimate the
intrinsic spectrum of the source.  From this estimate of the
intrinsic spectrum we computed the optical depth needed to
reproduce the observed flux of the remaining image by 
computing the necessary optical depth at 5~GHz, $\tau_5$, and 
its spectral index, $\alpha$, under the assumption that 
$\tau \propto \nu^\alpha$ between 5~GHz and the available higher
frequency.  We estimated the errors solely from the published
measurement errors, so they include no uncertainties arising from
the lens model or any time variability between the measurement
epochs for the different frequencies.  

\begin{figure}
\centerline{\psfig{figure=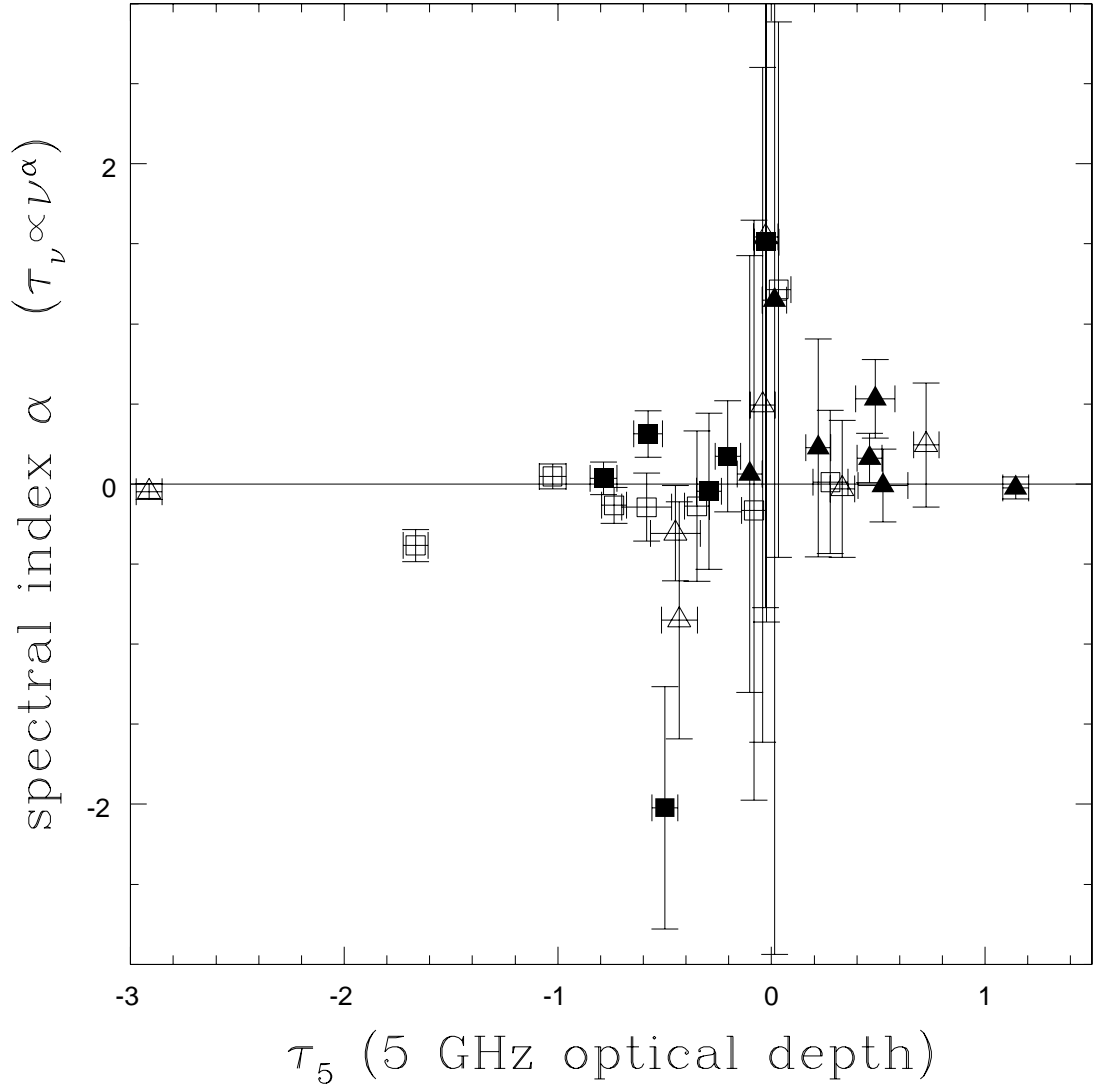,width=6.0in}}
\caption{Estimates of the ISM properties needed to explain the anomalous
  flux ratios.  We show an estimate of
  optical depth at 5~GHz, $\tau_5$, and the spectral index $\alpha$, 
  where $\tau \propto \nu^\alpha$, needed to bring the observed flux
  of each image into agreement with the average source properties predicted
  by the other three images.  The images are coded by their parities and
  relative fluxes with squares used for minima (positive parity) and    
  triangles for saddle points (negative parity).  Filled symbols are
  used for the brightest image of each parity, and open symbols for
  the faintest image.  Note the concentration of the brightest saddle
  points (minima) towards positive (negative) optical depth.
  }
\label{fig:freq}
\end{figure}

The results are shown in Fig.~\ref{fig:freq}.  We have coded the
images by their parities (minima versus saddle points) and 
relative fluxes (brightest versus faintest for each parity) 
because we will show below that the distributions of the model
flux residuals depend strongly on these image identifications.
This is seen in Fig.~\ref{fig:freq} as the concentration of the
bright saddle-point images in the direction of higher optical
depth compared to the other images.  We have dropped four (of 32)
images where the estimated optical depth changes sign between
the higher and lower frequencies and it is impossible to define a 
spectral index.  Except at zero optical depth, where the slope estimates 
become unstable and have larger uncertainties, the lenses broadly
require $\alpha \simeq 0$.  Since most radio propagation effects have 
strong wavelength scalings, the ISM seems an unlikely explanation.

\subsection{A Statistical Test for Substructure}
\label{sec:parity}

While the spectral indices required to explain the anomalous flux ratios
are peculiar for standard scattering or absorption processes, we would
prefer to have a test which can distinguish any ISM effect from the
gravitational effects of substructure.  One approach is to note that 
the physical properties of the ISM are a local property of the lens, and 
should have no knowledge of the global properties of the lens geometry.  
Radial gradients in the properties of the ISM should matter little because
in each lens the images of interest effectively lie at the same radius and there is no
simple angular correlation between the positions of the images and the 
major axis of the galaxy.  Moreover, creating the anomalous flux ratios
depends on the presence of clumped ISM components rather than smoothly
distributed components both to create physical conditions extreme enough
to have an effect and to differentially affect the lensed images.
The only exception to this rule arises for effects such as scintillation
or scatter broadening where the effect diminishes as the source
size becomes larger -- for these cases, the ISM should preferentially
perturb the least magnified images because they are the most compact.

The effects of gravity, however, are not determined purely by local
conditions because the image positions and fluxes 
are determined by the competition between the local gravitational potential 
and the geometric time delay.  The magnification tensor, whose eigenvalues
determine the image parities, depends on the projected surface density,
the projected tidal shear in the gravity, and the redshifts of the lens
and the source. As a result, we can attach a global identification to each 
image based on its parity (maximum, minimum, saddle point of the time delay 
surface) as well as its magnification with which the ISM should show
little (magnification) or no (parity) correlation.  Image parities 
and magnification orderings (but not the precise magnifications)
are generic predictions of successful lens models.  
In particular, for a 
four image lens, the images alternate parities in the sequence
minimum--saddle point--minimum--saddle point as we go around
the lens.  We can also identify the most and least magnified 
image of the two minima or saddle points from the geometry of
the lens.  Unlike the ISM, the effects of substructure depend on the image
parity and magnification.  The fluxes of highly magnified images
are more unstable to small gravitational perturbations than
the fluxes of less magnified images (as originally emphasized
by Mao \& Schneider~\cite{Mao98}). Most importantly, the
magnification perturbations created by low optical depth substructure 
have a very strong dependence on the image parity 
because the perturbations to the fluxes of the saddle 
points are skewed in the direction of demagnification
(Schechter \& Wambsganss~\cite{sw02}, Keeton~\cite{Keeton02}). 
This leads to a simple non-parametric test for distinguishing the
effects of the ISM from the effects of gravity -- {\it if the flux
anomalies depend on parity and magnification they cannot be due
to the ISM.}

\begin{figure}
\centerline{\psfig{figure=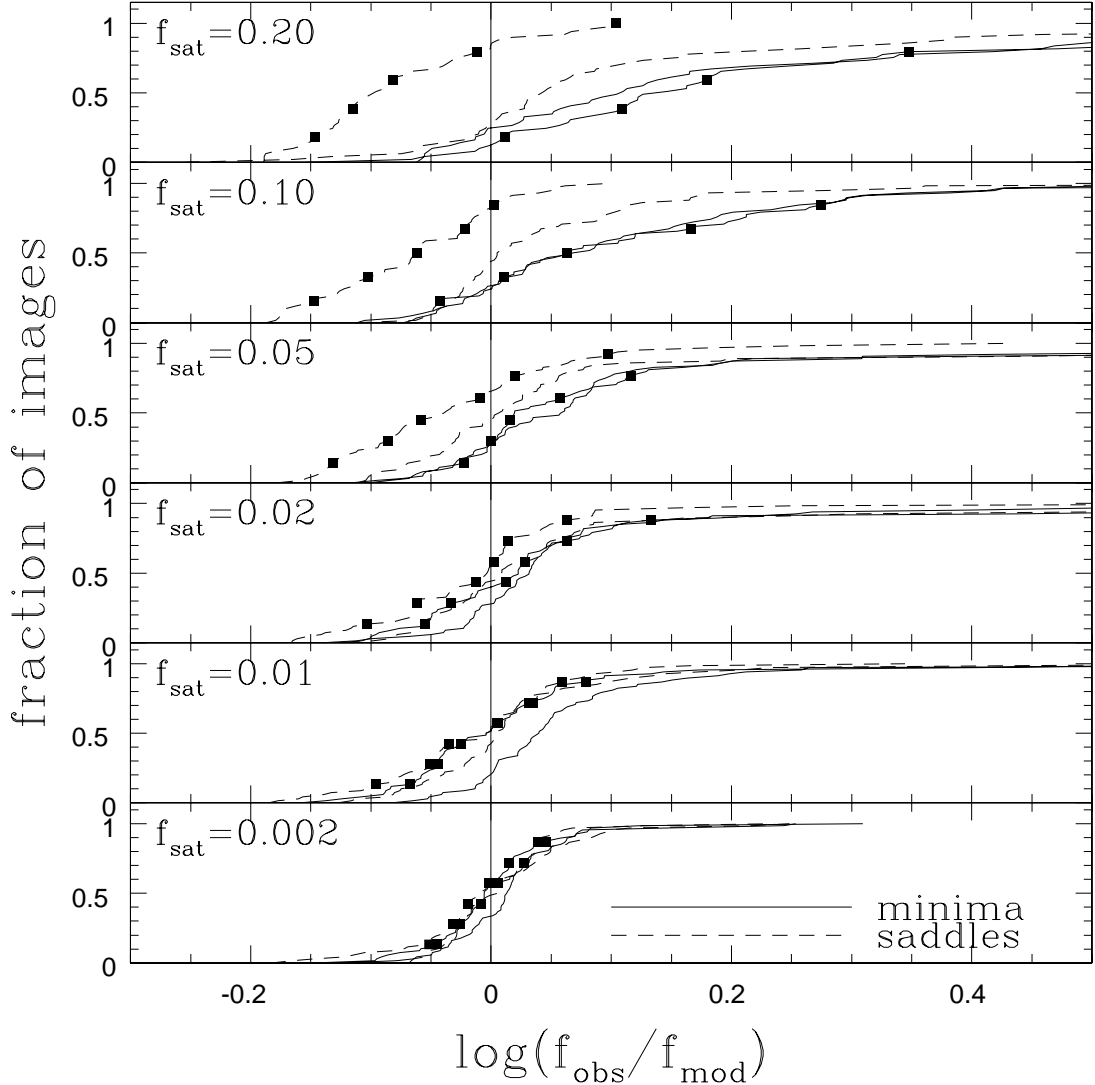,width=6.0in}}
\caption{Cumulative distributions of model flux residuals, $\log(f_{obs}/f_{mod})$,
  expected from Monte Carlo simulations of substructure with mass fraction
  $f_{sat}$ between $0.20$ (top) and $0.002$ (bottom).  The solid
  (dashed) lines are for minima (saddle points), with squares (no squares)
  for the distribution corresponding to the most (least) magnified image.
  Note the steady shift of the brightest saddle point to fainter
  flux residuals as the satellite fraction increases.  For very
  large satellite fractions the distributions would become more
  similar.
  The Monte Carlo samples are based on the same lens sample as was
  used in DK02 assuming 10\%
  flux measurement errors.
  \label{fluxresid:monte}
  }
\end{figure}

\begin{figure}
\centerline{\psfig{figure=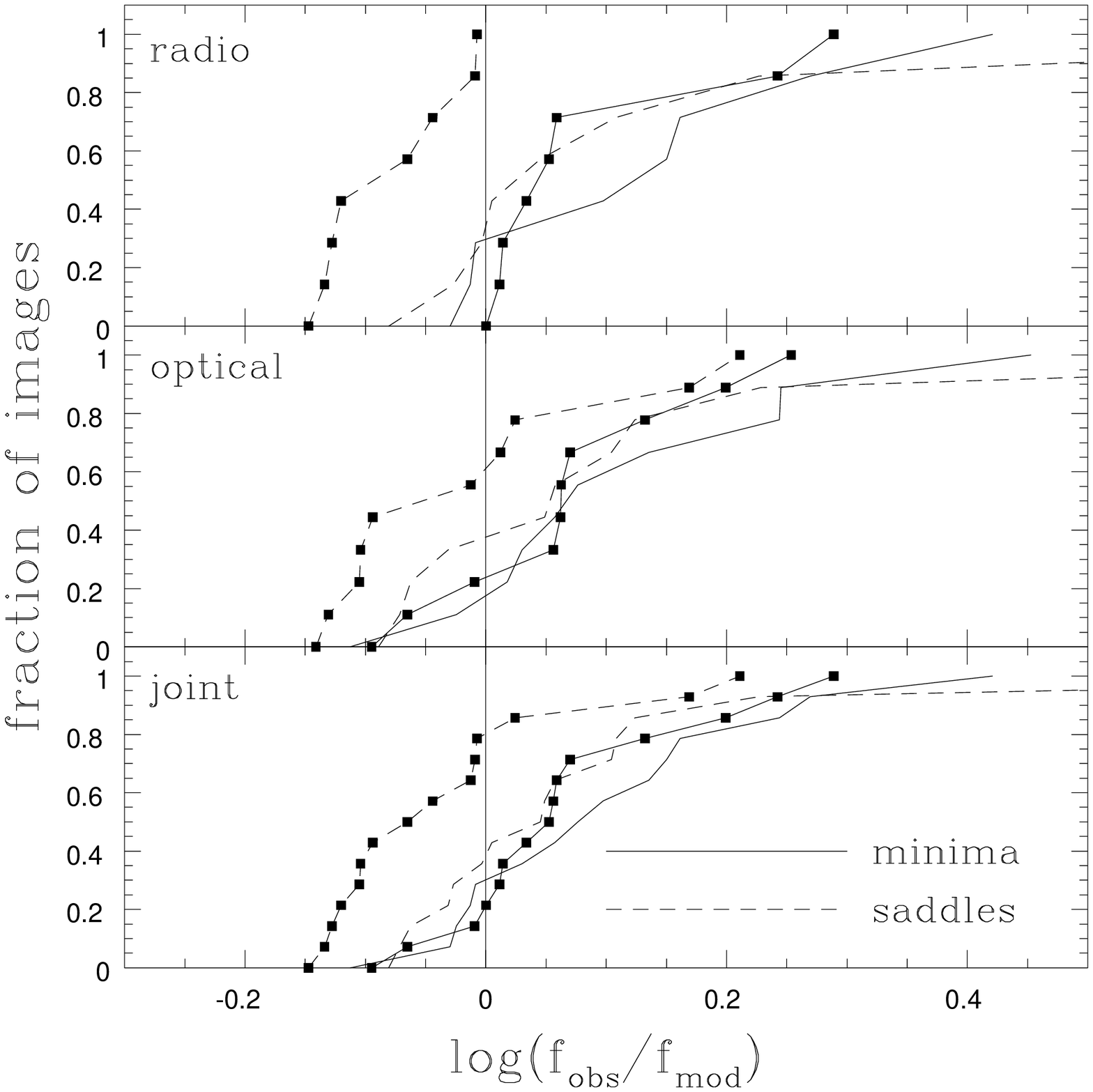,width=6.0in}}
\caption{Cumulative distributions of model flux residuals, $\log(f_{obs}/f_{mod})$,
  in the real data, assuming constant fractional flux errors for each image.  The solid
  (dashed) lines are for minima (saddle points), with squares (no squares)
  for the distribution corresponding to the most (least) magnified image.
  From top to bottom the distributions are shown for samples of 8 radio,
  10 optical or 15 total four-image lenses.  If the flux residuals are
  created by propagation effects we would not expect the distributions
  to depend on the image parity or magnification, while if they are due
  to low optical depth substructure we would expect the distribution for
  the brightest saddle points to be shifted to lower observed fluxes.
  Note the similarity between the observed curves and the Monte Carlo
  simulations for substructure fractions of order $\sim 5$\%.
  \label{fluxresid:data}
  }
\end{figure}

We first demonstrate the effects of substructure on the distribution
of flux residuals.  We took the same sample of 7 lenses we used
in DK02 (MG0414+0534, Hewitt et al.~\cite{Hewitt92}; 
B0712+472, Jackson et al.~\cite{Jackson98}; PG1115+080, Weymann et al.~\cite{Weymann80};
B1422+231, Patnaik et al.~\cite{Patnaik92}; B1608+656, Fassnacht et al.~\cite{Fassnacht96};
B1933+503, Sykes et al.~\cite{Sykes98}; and B2045+265, Fassnacht et al.~\cite{Fassnacht99}). 
The data were fit using standard
macro models (singular isothermal ellipsoids (SIE) plus external shear) 
to generate an initial model.  We then added a substructure
mass fraction $f_{sat}$ near the images, modeling the substructure
as tidally truncated isothermal spheres with critical radius $b=0\farcs001$
(corresponding to approximately $10^6 M_\odot$).
With the substructure added we reidentified the lensed images and
their fluxes and added astrometry ($0\farcs003$ rms) and flux (10\% rms)
measurement errors to generate a model data set with substructure.
This model data was then refit using the standard macro models,
including no weight on the fit to the fluxes of the images.
We then compare the observed image fluxes, $f_{obs,i}$ to the model 
image fluxes, $f_{mod,i}=|M_i| f_{src}$, found given the 
magnification $M_i$ predicted by the model fitted to the perturbed
data.  We generated 10 such
realizations for each of the 7 lenses.  Figure~\ref{fluxresid:monte}
shows the distributions of the residuals, $\log(f_{obs,i}/f_{mod,i})$,
for the individual image types (saddle point/minimum, brightest/faintest)
as a function of the substructure mass fraction, $f_{sat}$.
As expected, the residual distribution for the brightest saddle
point is markedly different from that for all other images unless
the substructure fraction is so low as to be masked by the 
random flux errors.

We now repeat the test for the real data. The same standard macro
models were used for each system, and we again assigned no weight
to fitting the observed image fluxes.  For radio lenses we used the highest
frequency measurements from either the VLA or Merlin, and for the
optical lenses we used the flux ratios at the longest available 
wavelength (generally H-band where none of the systems shows
significant extinction).  This gave us a sample of 8 radio systems
(adding B0128+437, Phillips et al.~\cite{Phillips00}, and B1555+375,
Marlow et al.~\cite{Marlow99b} to the DK02 sample), 10 optical systems
\footnote{The systems are the four-image quasar
  lenses HE0230--2130 (Wisotzki et al.~\cite{Wisotzki99}),
  HE0435--1223  (Wisotzki et al.~\cite{Wisotzki02}),
  HS0810+2554 (Reimers et al.~\cite{Reimers02}),
  RXJ0911+0551 (Bade et al.~\cite{Bade97}),
  PG1115+080 (Weymann et al.~\cite{Weymann80}),
  H1413+117 (Magain et al.~\cite{Magain88}),
  Q2237+030 (Huchra et al.~\cite{Huchra85}),
  B1422+231 (Patnaik et al.~\cite{Patnaik92}),
  B2045+265 (Fassnacht et al.~\cite{Fassnacht99}) and
  MG0414+0534 (Hewitt et al.~\cite{Hewitt92}).  
  The latter 3 systems are also in the radio sample.
  We used the Castles H-band fluxes if available, otherwise the
  I-band fluxes.
   }
and 15 joint systems.  
For our standard model we estimated the
source flux using constant fractional errors on the image fluxes.
Figure~\ref{fluxresid:data}
shows the distribution for the residuals, $\log(f_{obs}/f_{mod})$
for all presently available quads containing either compact radio sources
or bright quasars.  Several systems appear in both categories, and
for these systems we use the radio fluxes for the joint sample. 
The distributions of the residuals differ for
minima and saddles in the sense that saddle points tend to be
fainter than expected and minima tend to be brighter -- just
as in the simulations of the distributions expected for
substructure in Fig.~\ref{fluxresid:monte}.   

We used the Kolmogorov-Smirnov (KS) test 
to estimate the significance of the differences.  If we
simply compare the distributions of $\log(f_{obs}/f_{mod})$ for 
minima and saddle points, we find that the KS test probabilities for
the two distributions to be the same are 2.3\%, 28\%, 5.5\% for
the radio, optical and joint samples respectively.  Theoretically,
we expect the most highly magnified saddle point to have the 
most discrepant residual distribution because the high magnification
makes the flux unstable to perturbations and because the theoretical
studies of low optical depth substructure predict that it should
show the largest differences (as in Fig.~\ref{fluxresid:monte}).  
The KS test probabilities
for the most magnified saddle point to have the same residual
distribution as the other three images is only 0.04\%, 5\% and
0.3\% for the radio, optical and joint samples.    
In each case, the most highly
magnified minimum shows the next lowest probability for a residual
distribution agreeing with the other images (1.9\%, 59\% and 18\%
respectively), but the significance is lower.  In the case
of substructure, the magnification perturbation distribution for
the minima is harder to differentiate from the measurement uncertainties
which probably dominated the residuals for the faintest images.

We can check these significance estimates by examining trials in
which the identifications of the images are randomized.  We made
$10^4$ trials in which we assigned the image identifications 
(1=brightest minimum, 2=faintest minimum, 3=brightest
saddle point, 4=faintest saddle point) to the images at random
with the restriction that each lens always have the 4 possible
image types.  We then estimated the probability that these
random assignments could produce results similar to that 
found for the true assignments.  The fraction of the time the
observed values of the KS statistic were exceeded tracked the
KS test probability estimates well.  For example, in only 
2.3\%, 18\% and 4.8\% of the trials did the KS statistic
for the comparison between the distributions of minima
and saddle points exceed that of the true distribution.
We also examined the distribution of the KS statistic if we
bootstrap resampled the lens sample. The KS statistics and
probabilities were typical of any sample of lenses drawn 
(with replacement) from the available 4-image lenses, so
the results are not dependent on any particular system or
small subset of systems.

We also examined weighting schemes for estimating the
source flux other than assuming constant fractional errors
for the observed image fluxes.  Note, however, that our
test is independent of any simple rescaling of the source
flux.  The KS test makes a non-parametric comparison of
the  $\log(f_{obs}/f_{mod})$ distributions, and a uniform
change in the source flux simply shifts the distribution.  If
we assume constant flux errors rather than constant fractional
errors, the probability the saddles and minima have the same
distribution is again 2.3\%. If we set the source flux using
the fluxes of the faintest minimum and saddle, then the 
probability that the brightest minimum and saddle have the
same residual distribution for the radio lenses is 0.02\%.  

Finally, we can compare the observed distribution to either those
expected for Gaussian errors or those we calculated for our
Monte Carlo simulations of substructure.  For example, while
the overall residual distributions of the images are well fit
by a log-normal distribution of width $\sigma_{\ln f}=0.24$,
the residual distributions of the images separately are 
grossly inconsistent with log-normal distributions.  The
KS test likelihood for the joint distribution of all images
matching a log-normal distribution is $0.26$, while the
likelihood for the distributions of the individual types 
matching a single log-normal distribution is $1.5\times 10^{-5}$.
This is simply another way of quantifying the fact that
the residual distributions depend on the image parities
and flux orderings.  If we compare the residual distributions
by image for the radio lens sample to the distributions 
predicted in the Monte Carlo simulations, we find KS
test probabilities for the distributions to be the same
of $4 \times 10^{-7}$, $2 \times 10^{-5}$, $2\times 10^{-4}$,
$0.02$, $0.3$ and $0.05$ for the cases with
satellite mass fractions of $f_{sat}=0.002$, $0.01$, $0.02$, 
$0.05$, $0.10$ and $0.20$ respectively.  While we do not,
at present, feel this is as optimal a way of estimating
$f_{sat}$ as the methods we used in Dalal \& Kochanek~ (\cite{dk02}),
it emphasizes the consistency of the observed dependence
of the residual distribution on parity and flux with that
expected for substructure.

\begin{figure}
\centerline{\psfig{figure=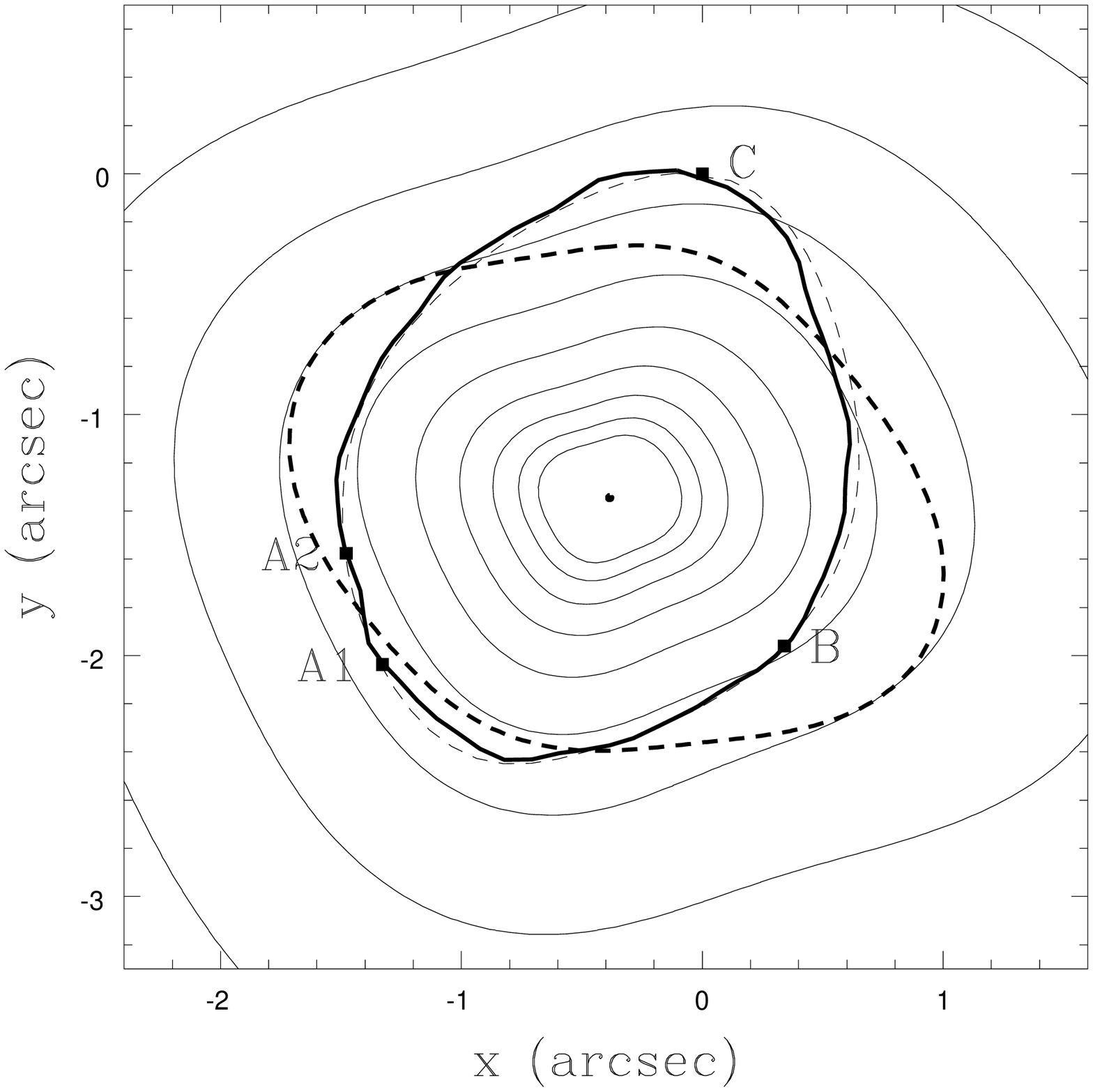,width=2.8in}
            \psfig{figure=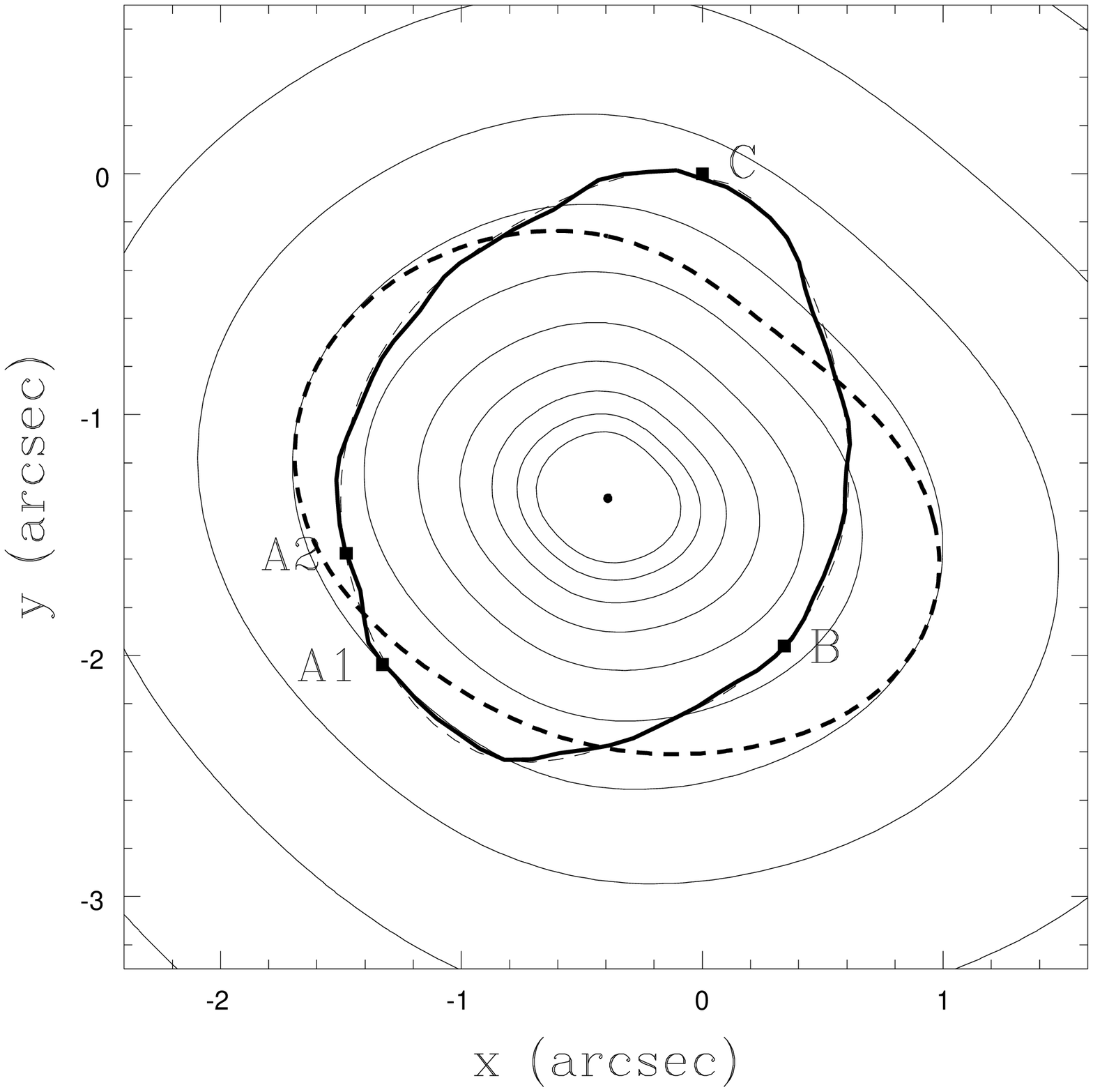,width=2.8in}}
\caption{Surface density contours for models of PG1115+080 including misaligned
  $a_3$ and $a_4$ multipoles (thin lines).  The model in the left panel is
  constrained only by the 4 compact images (A$_1$, A$_2$, B and C, filled
  squares).  The model in the right panel is also constrained by the Einstein
  ring image of the host galaxy.  The heavy dashed curve is the
  tangential critical curve of the lens.  The heavy solid curve is the
  ring curve of the host galaxy and the thin dashed curve is the best fit 
  model.
   }
\label{fig:pg1115}
\centerline{\psfig{figure=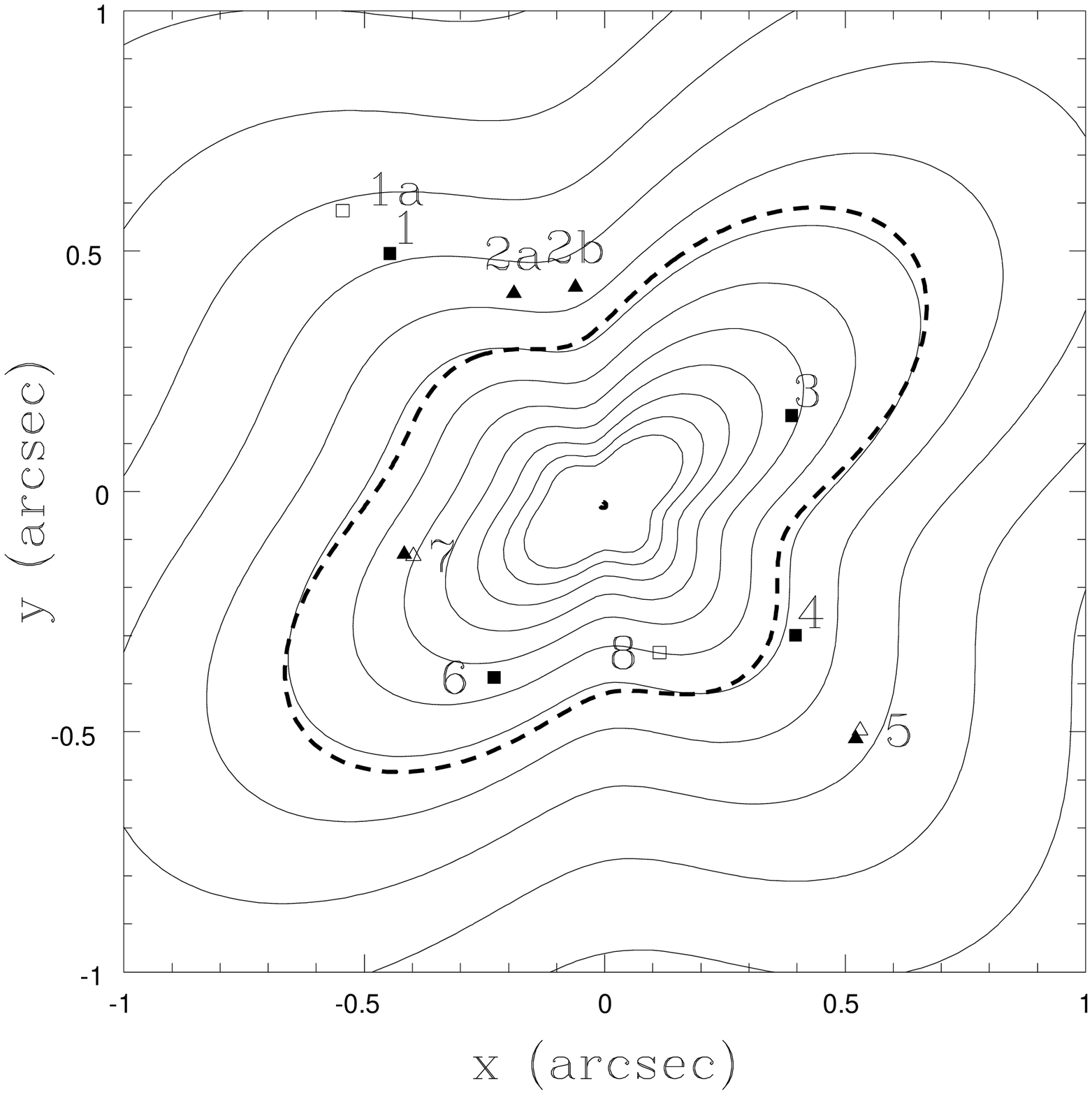,width=2.8in}
            \psfig{figure=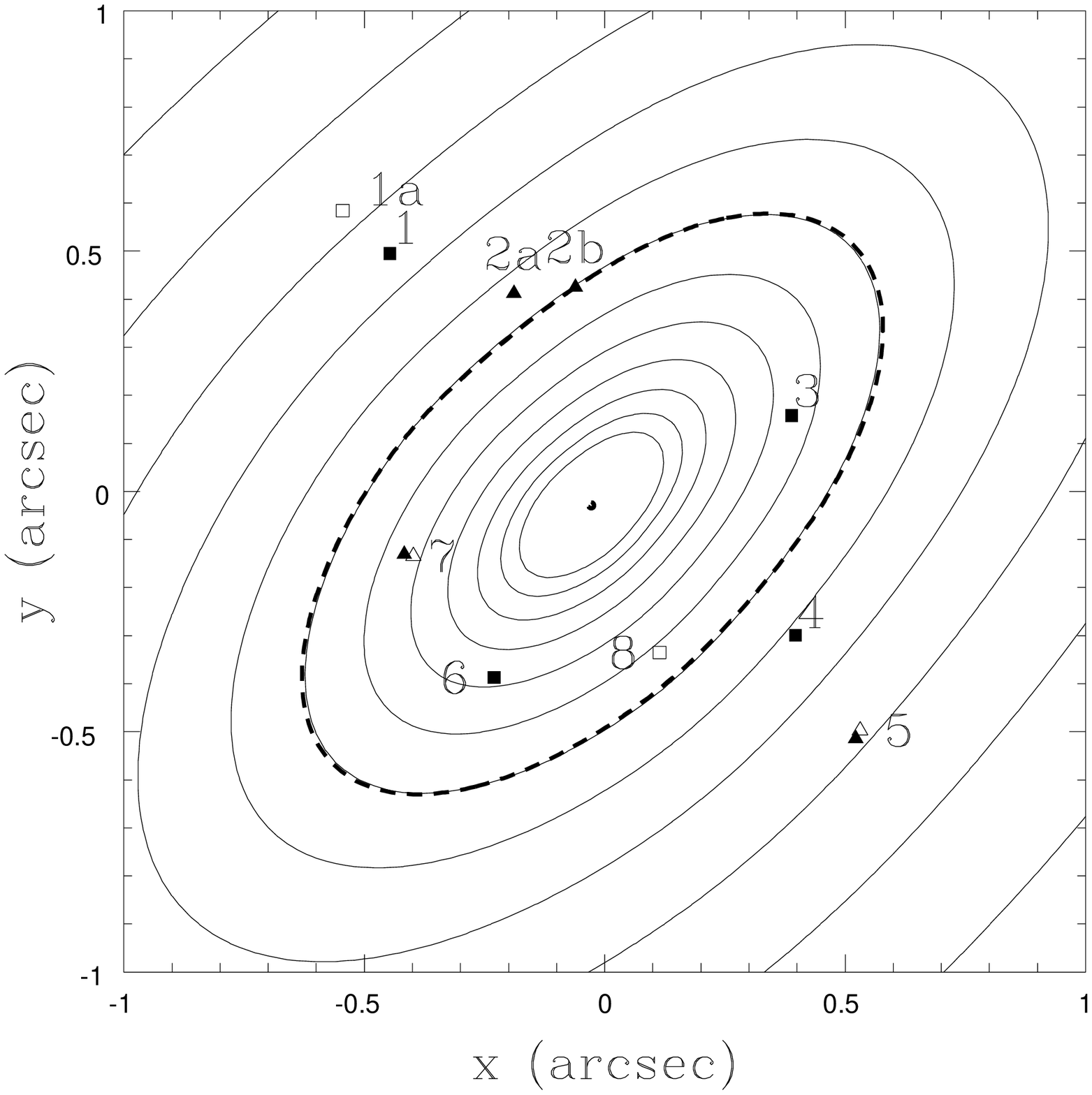,width=2.8in}}
\caption{Surface density contours for models of B1933+503 including misaligned
  $a_3$ and $a_4$ multipoles (thin lines).  The model in the left panel is
  constrained only by the 4 compact images (images 1, 3, 4 and 6, filled
  squares).  The model in the right panel is also constrained by the other 
  images in the lens (the two-image system 1a/8, open squares; the four-image system
  2a/2b/5/7 filled triangles; and the two-image system comprising parts of 5/7,
  open triangles)   The tangential critical line of the model (heavy solid
  curve) must pass between the merging images 2a/2b, but fails to do so in
  the first model (left panel).     
   }
\label{fig:b1933}
\end{figure}

\begin{figure}
\centerline{\psfig{figure=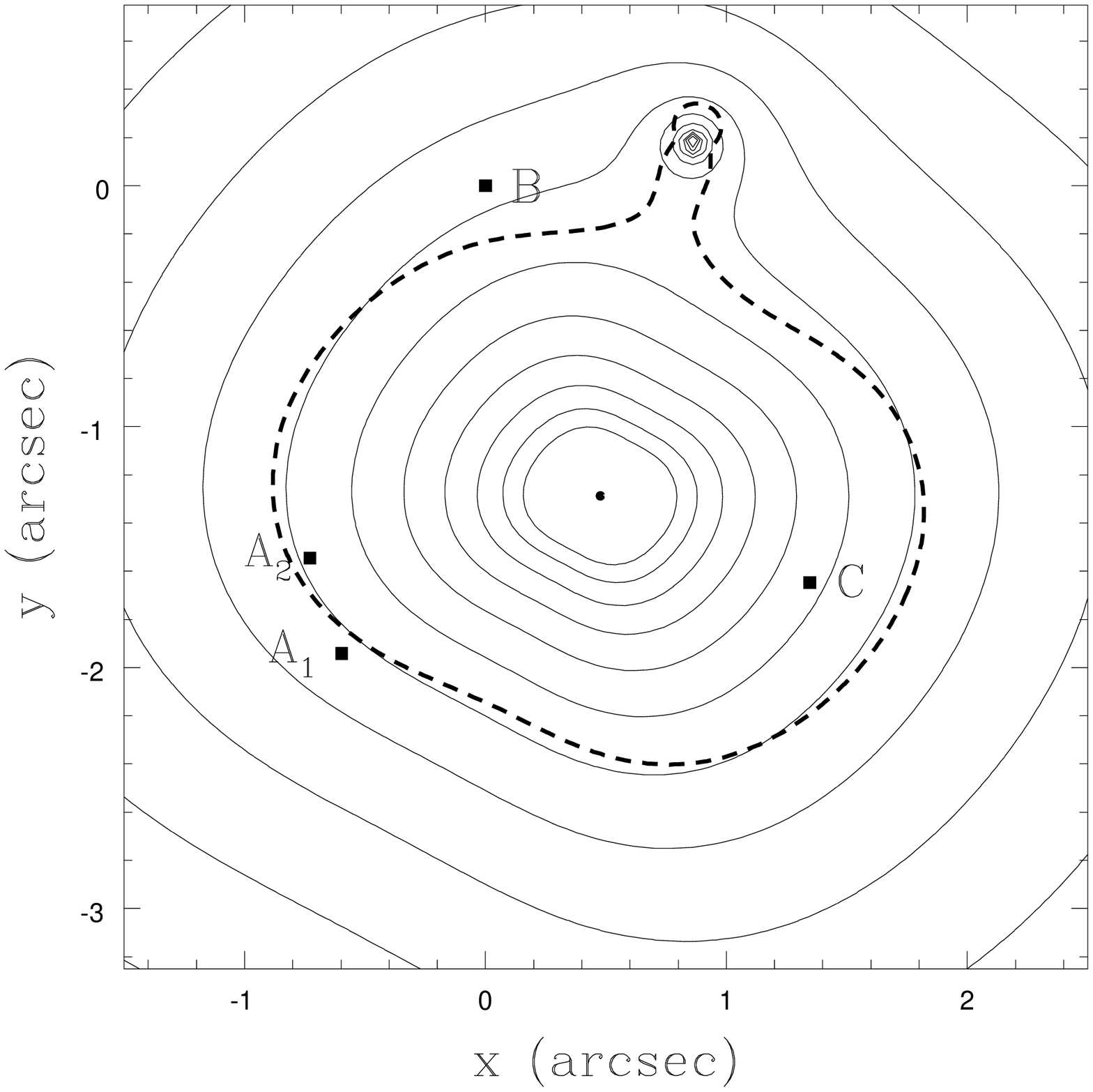,width=2.8in}
            \psfig{figure=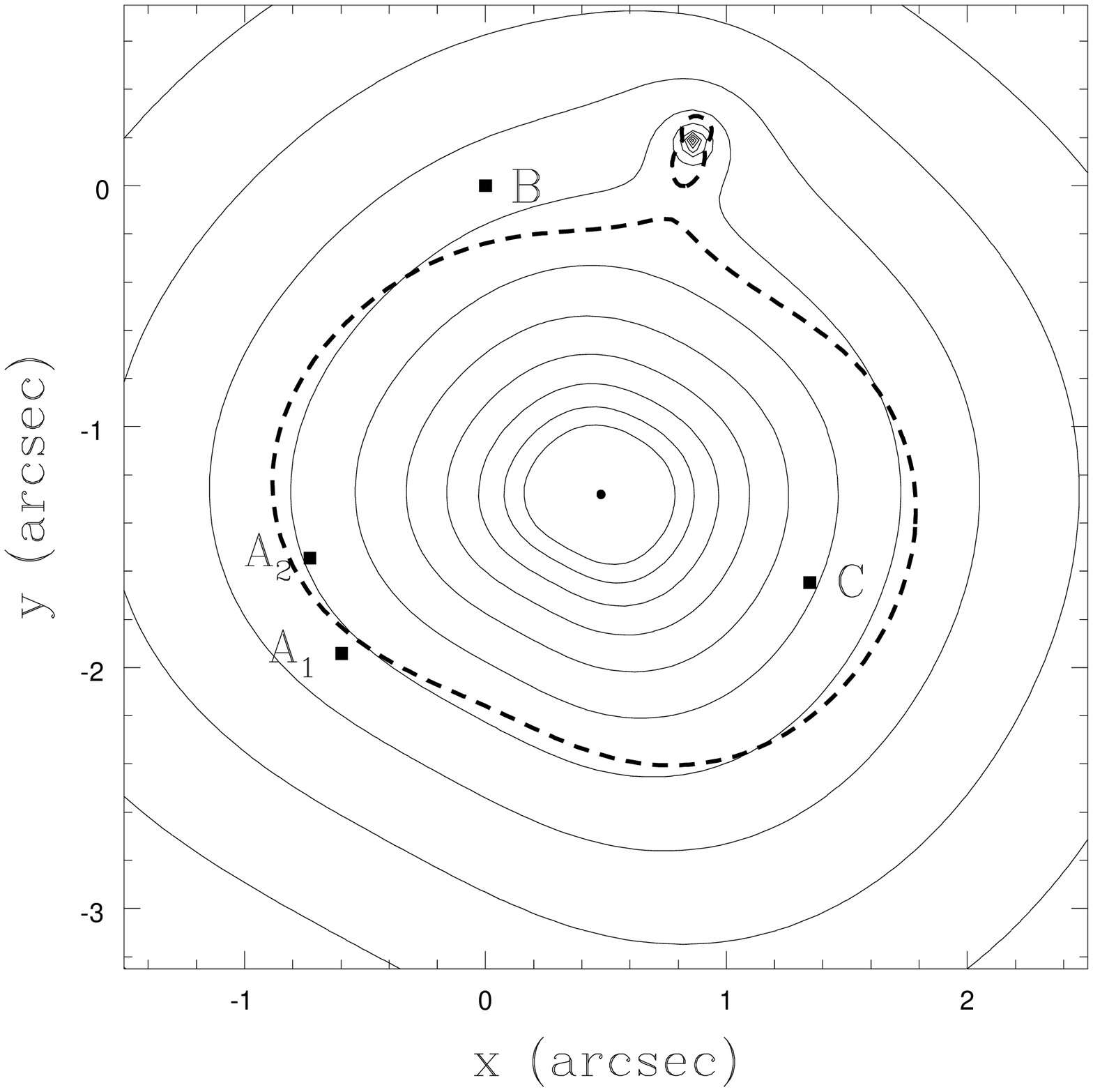,width=2.8in}}
\caption{Surface density contours for models of MG0414+0534 including misaligned
  $a_3$ and $a_4$ multipoles (thin lines).  The model in the left panel is
  constrained only by the 4 compact images (A$_1$, A$_2$, B and C, filled
  squares).  The model in the right panel is also constrained by the structure
  of the VLBI jets associated with each image.  The heavy dashed curve is the
  tangential critical curve of the lens.  Note the small satellite galaxy to
  the North of the lens -- this is an example of the high mass substructures
  ($M \sim 10^{10} M_\odot$)
  which were included as part of the macro model in DK02 rather than treated
  as substructure.
   }
\label{fig:mg0414}
\end{figure}

\section{Ruling Out Systematic Problems in the Model Potentials}

If the anomalous flux ratios cannot be explained by propagation effects
in the ISM, the only alternative to some form of substructure in the
gravitational field is a systematic problem in the assumed macro model
for the lens galaxy.  Previous studies of the effects of changes in
the macro model show that changes in the radial structure from the
standard singular isothermal ellipsoids (SIE) with external shear
do not provide a solution (Metcalf and Zhao \cite{mz02}, 
Keeton et al.~\cite{Keetonetal02}). This makes physical sense because
the images in most of the systems with anomalous flux ratios all lie
close to the Einstein radius of the lens, making it difficult for changes in the
radial profile to produce changes in the flux ratios. We focus on
changes in the angular structure because they can more easily shift
the positions of the tangential critical line so as to change the
image fluxes.

\subsection{Lens Models With Deviations from Ellipsoidal Structure}

We modified the gravitational potential by adding higher order multipoles
of the form
\begin{equation}
   \phi_m   =  { \epsilon_m \over m } r \cos m(\theta-\theta_m)
\end{equation}
to the primary lens galaxy.  These correspond to a surface density
\begin{equation}
   \kappa_m =  { \epsilon_m \over r } { 1- m^2 \over m} \cos m(\theta-\theta_m)
\end{equation}
and match the expansion used by Evans \& Witt~(\cite{Evans02}).  The new terms
have been included in a revised version of the {\it lensmodel}
package (Keeton~\cite{Keeton01b}).\footnote{The {\it lensmodel} package
now allows terms of the form $(\epsilon_m/m) r^n \cos m(\theta-\theta_m)$ where the
normalization was chosen to reduce to an external shear for $n=2$
and $m=2$.}  In {\it lensmodel}, the surface density of the SIE model
is defined by
\begin{equation}
   \kappa = { 1 \over \sqrt{2} } { b \over r }
    { q \over \left[ 1+q^2 - (1-q^2) \cos 2(\theta-\theta_2)\right]^{1/2} }
\end{equation}
where $q=1-\epsilon_2$ is the axis ratio and $\theta_2$ and is orientation of
the major axis.  The critical line, which is an isodensity contour with
$\kappa=1/2$ for the SIE, has a major axis of $b$ and a minor axis of $bq$.
Thus, if we add an $m=4$ term to an SIE model, these normalizations imply
a standard amplitude for the deviation of an isophote from the ellipse
of $a_m = \epsilon_m (1-m^2)/ m b$.  Positive $\epsilon_4$ corresponds to 
positive $a_4$ and ``disky'' density contours when the $m=4$ component is
aligned with the quadrupole ($\theta_4 = \theta_2$).  Galaxies with
``boxy'' isophotes will have negative $a_4$ and $\epsilon_4$. 

We will restrict our attention to adding $m=3$ and $m=4$ terms to the
potential of the primary lens galaxy.  This should be sufficient to
indicate their potential importance without the models being too
under-constrained.  Observational evidence indicates that
the amplitude of these higher order multipoles is small.
For example, Bender et al.~(\cite{bender}) and Rest et al.~(\cite{Rest01})
examine
a sample of local ellipticals and quantify the deviations of the
isophotes from ellipses.  They find that the deviations are dominated
by the $m=4$ term, $\propto a_4 \cos(\theta-\theta_4)$, with a typical amplitude
of $|a_4|\sim 0.01$.  Deviations from ellipticity in projections
of simulated halos are of comparable amplitude
(Heyl et al.~\cite{heyl}, Burkert \& Naab~\cite{Burkert03}).  
Considerably less is available on the
distribution of misalignment angles, $\theta_m-\theta_2$, between the 
higher order multipoles and the dominant ellipse.
We know from the statistics of lens models that the
major axes of the mass and the light are closely aligned 
(Kochanek~\cite{Kochanek01b}).
The effect of these higher order terms on
lenses were first studied by Evans \& Witt~(\cite{ew}), who pointed out that
larger values of $a_4$ than are typically observed would
lead to significant cross sections for lenses producing more than 4
observable images.  

We fit the 7 lenses in the DK02 sample with a sequence of four models
whose results are presented in Table~\ref{tab:mod1}.  In the first model 
(``astrometry'') we fit only the image positions using our standard SIE
plus external shear model.  In the second model (``astrometry$+$flux'')
we added the constraints from the image fluxes assuming 5\% uncertainties.  
In either case, 6 of the 7 lenses have unacceptably high $\chi^2_{flx}$ 
fit statistics for the image fluxes even though the standard models provide
good fits to the astrometric data, $\chi^2_{tot}-\chi^2_{flx}$.  
Only B1608+656 has a $\chi_{flx}^2$ largely consistent with 5\% measurement 
uncertainties.   The failure of these models to reproduce the flux ratios is the 
origin of the entire problem of flux ratio anomalies.  
In the third model we added the
$m=3$ and $m=4$ multipoles to the potential of the primary lens, but
constrained them to be aligned with the ellipsoid ($\theta_m=\theta_2$
but allowing any sign for $\epsilon_m$). For MG0414+0534, B0712+472,
PG1115+080 and B1933+503 the new terms lead to a significant reduction
in $\chi^2_{flx}$, but in no case does it become a statistically
acceptable fit.  The amplitudes of the multipoles are unreasonably
large for B0712+472 and B1933+503.  Finally, in the fourth model, we
allowed the $m=3$ and $m=4$ multipoles to be misaligned with respect
to the ellipsoid.  In these models the flux ratio anomalies implied
by the standard lens models for PG1115+080, B1422+231 and B1933+503
are gone, but the multipole amplitudes are still unreasonably large.

Thus, like Evans \& Witt~(\cite{Evans02}), we find that adding higher
order multipoles to the potential could explain some of the flux ratio
anomalies (3 of the 6 in the DK02 sample), if the positions
and fluxes of the 4 compact images were the only available data.  The surface density 
contours implied by these models are not very physical, as shown
for PG1115+080 and B1933+503 in Figs.~\ref{fig:pg1115} and 
\ref{fig:b1933}.  The improved fit with the higher order multipoles
is easily understood for B1933+503.  In this lens, image 4 is 
anomalously bright for an ellipsoidal model, and the new models
solve the problem by making the tangential critical line lie
much closer to image 4 so that it can be more highly magnified.
The cause of the improvement in PG1115+080, where the anomaly is 
due to the flux ratio between images A$_1$ and A$_2$, is less
obvious.  In other cases, like MG0414+0534, the surface density 
contours, while very boxy, look more reasonable (Fig.~\ref{fig:mg0414}).

For three of the DK02 lenses with anomalies there are additional lensed components 
beyond the four compact images which can be used to test these models.
In MG0414+0534 there are VLBI jets associated with each of the quasar
images (Ros et al.~\cite{Ros00}, Trotter, Winn \& Hewitt~\cite{Trotter00}),
in PG1115+080 there is the Einstein ring image of the quasar
host galaxy (Impey et al.~\cite{impey98}, Kochanek et al.~\cite{Kochanek01}),
and in B1933+503 there are additional multiply-imaged components of the radio
source (Sykes et al.~\cite{Sykes98}, Cohn et al.~\cite{Cohn01},
Munoz, Kochanek \& Keeton~\cite{munoz01}).  Table~\ref{tab:mod2}
shows the results of fitting the same series of models to these
three lenses with the addition of the new constraints. For MG0414+0534,
where the models were already unable to explain the anomalous flux
ratios, little changes with the addition of the VLBI constraints,
although the amplitudes of the multipoles change as shown in
Fig.~\ref{fig:mg0414}.  For the two lenses where it appeared possible
for the higher order multipoles to eliminate the anomalies, we find
that the new constraints eliminate these solutions.    
Figs.~\ref{fig:pg1115} and
\ref{fig:b1933} show the surface density contours for these new
models, and it is immediately clear that the added constraints
force the lenses to be significantly more regular and ellipsoidal.

\begin{deluxetable}{crrrrrrrl}
\tablecaption{Basic Model Fits }
\tablewidth{0pt}
\tablehead{
  Lens &Case &$N_{dof}$ &$\chi^2_{tot}-\chi^2_{flx}$ &$\chi^2_{flx}$  &$|a_3|$  &$|a_4|$ }
\startdata
MG0414+0534 &astrometry                &$2$ &$ 12.8$ &$ 41.2$ \nl
            &astrometry+flux           &$5$ &$ 10.0$ &$ 35.5$ \nl
            &$\delta\theta_m\equiv 0$  &$3$ &$  2.4$ &$ 19.1$ &$0.011$ &$0.020$ \nl
            &full                      &$1$ &$  6.3$ &$ 14.4$ &$0.005$ &$0.040$ \nl
\hline
B0712+472   &astrometry                &$1$ &$  4.2$ &$112.7$ \nl
            &astrometry+flux           &$4$ &$ 31.8$ &$ 15.5$ \nl
            &$\delta\theta_m\equiv 0$  &$2$ &$  6.6$ &$ 21.4$ &$0.095$ &$0.178$ \nl
            &full                      &$0$ &$  1.9$ &$ 17.2$ &$0.242$ &$0.416$ \nl
\hline
PG1115+080  &astrometry                &$2$ &$  2.9$ &$ 57.0$ \nl
            &astrometry+flux           &$5$ &$ 10.0$ &$ 40.3$ \nl
            &$\delta\theta_m\equiv 0$  &$3$ &$  2.2$ &$ 18.3$ &$0.006$ &$0.022$ \nl
            &full                      &$1$ &$  2.9$ &$  3.7$ &$0.022$ &$0.051$ \nl
\hline
B1422+231   &astrometry                &$1$ &$  2.1$ &$ 12.3$ \nl
            &astrometry+flux           &$4$ &$  2.7$ &$ 11.2$ \nl
            &$\delta\theta_m\equiv 0$  &$2$ &$  2.7$ &$ 11.3$ &$0.000$ &$0.000$ \nl
            &full                      &$0$ &$  0.5$ &$  0.1$ &$0.089$ &$0.069$ \nl
\hline
B1608+656   &astrometry                &$2$ &$  1.8$ &$  3.9$ \nl
            &astrometry+flux           &$5$ &$  1.6$ &$  0.7$ \nl
            &$\delta\theta_m\equiv 0$  &$3$ &$  2.2$ &$  0.5$ &$0.000$ &$0.000$ \nl
            &full                      &$1$ &$  4.1$ &$  0.6$ &$0.000$ &$0.000$ \nl
\hline
B1933+503   &astrometry                &$1$ &$  0.5$ &$254.2$ \nl
            &astrometry+flux           &$4$ &$ 19.2$ &$137.4$ \nl
            &$\delta\theta_m\equiv 0$  &$2$ &$  1.6$ &$  0.1$ &$0.011$ &$0.137$ \nl
            &full                      &$0$ &$  1.3$ &$  0.0$ &$0.079$ &$0.138$ \nl
\hline
B2045+265   &astrometry                &$1$ &$  1.4$ &$302.0$ \nl
            &astrometry+flux           &$4$ &$  2.1$ &$301.4$ \nl
            &$\delta\theta_m\equiv 0$  &$2$ &$  2.2$ &$301.4$ &$0.000$ &$0.000$ \nl
            &full                      &$0$ &$  1.9$ &$301.3$ &$0.001$ &$0.000$ \nl
\enddata
\tablecomments{
  Four model cases are fit: ``astrometry'' and ``astrometry$+$flux'' fit standard models
  (SIE $+$ shear) to the image positions and fluxes.  The ``$\delta\theta_m\equiv 0$''
  adds the $m=3$ and $m=4$ multipoles constrained to be aligned with the ellipsoid
  ($\theta_m=\theta_2$), and the ``full'' models allow any orientation.  We show the
  number of degrees of freedom $N_{dof}$, the goodness of fit 
  $\chi^2_{tot}-\chi^2_{flx}$ for the constraints other than the flux ratios
  and the goodness of fit to the fluxes $\chi^2_{flx}$ assuming 5\% errors in the
  fluxes.  The amplitudes of the multipoles are given by $|a_3|$ and $|a_4|$.  Very
  weak priors are included on the ellipticity and shear.
  }
\label{tab:mod1}
\end{deluxetable}

\begin{deluxetable}{crrrrrrrl}
\tablecaption{Fits Testing the Basic Model}
\tablewidth{0pt}
\tablehead{
 Lens &Case &$N_{dof}$ &$\chi^2_{tot}-\chi^2_{flx}$ &$\chi^2_{flx}$  &$|a_3|$  &$|a_4|$ }
\startdata
MG0414+0534 &astrometry                &$20$ &$ 52.7$ &$ 20.8$ \nl
            &astrometry+flux           &$23$ &$ 53.0$ &$ 19.6$ \nl
            &$\delta\theta_m\equiv 0$  &$21$ &$ 19.5$ &$ 17.6$ &$0.000$ &$0.032$ \nl
            &full                      &$19$ &$  6.3$ &$ 15.2$ &$0.021$ &$0.027$ \nl
\hline
PG1115+080  &astrometry                &$73$ &$222.1$ &$57.0$ \nl
            &astrometry+flux           &$76$ &$222.7$ &$55.6$ \nl
            &$\delta\theta_m\equiv 0$  &$74$ &$146.0$ &$13.8$ &$0.000$ &$0.027$ \nl
            &full                      &$72$ &$121.1$ &$22.9$ &$0.013$ &$0.021$ \nl
\hline
B1933+503   &astrometry               &$11$ &$ 34.4$ &$268.2$ \nl
            &astrometry+flux          &$14$ &$ 33.2$ &$256.1$ \nl
            &$\delta\theta_m\equiv 0$ &$12$ &$ 44.7$ &$231.2$ &$0.007$ &$0.001$ \nl
            &full                     &$10$ &$ 33.5$ &$242.5$ &$0.009$ &$0.000$ \nl
\enddata
\tablecomments{
  Model sequences including higher order multipoles for the three lenses with 
  flux ratio anomalies and additional
  model constraints.  The columns are the same as in Table~\protect{\ref{tab:mod1}}.
  }
\label{tab:mod2}
\end{deluxetable}

\begin{deluxetable}{lccccc}
\tablecaption{Compton Limits on CLASS Source Sizes}
\tablewidth{0pt}
\tablehead{
  Lens & $S$[mJy] & $\nu$ [GHz] & $z$ & $\mu_{\rm mod}$ &
   $\Delta\theta_{\rm min}$[$\mu$as] }
\startdata
B0128+437 & 18.9 & 5 & - & 4.3 & 22 \\
MG0414+0534 & 149 & 5 & 2.64 & 18 & 33 \\
B0712+472 & 10.5 & 5 & 1.33 & 17 & 7 \\
B1422+231 & 164 & 8.4 & 3.62 & 8.2 & 34 \\
B1555+375 & 17 & 5 & - & 5.2 & 18.6 \\
B1608+656 & 34.1 & 5 & 1.39 & 5.1 & 23.8 \\
B1933+503 & 17.6 & 8.4 & 2.62 & 3.7 & 14.7 \\
B2045+265 & 29.02 & 1.4 & 1.28 & 52 & 24 \\
\enddata
\tablecomments{Minimum source sizes for radio lenses, based on the brightest image.
  Sources without redshifts were assumed to have $z=2$.  The listed magnifications are
  from fits to the standard SIE + shear model. We assumed a brightness 
  temperature limit of $T_b=0.5 \times 10^{12}$~K.}
\label{compton}
\end{deluxetable}

\subsection{Statistical Checks of the Macro Model }

Thus, we find that higher order multipoles can explain only 1 (B1422+231)
of the 6 anomalous flux ratio lenses in the DK02 sample given the available data.  
Keeton et al.~(\cite{Keetonetal02}) also noted that B1422+231 was less
problematic than many of the other systems.  Nonetheless, the perturbation
amplitudes needed to explain B1422+231 are much larger than observed in
real galaxies, halo simulations or in the lenses where additional 
constraints allow us to determine $a_3$ and $a_4$.
We conclude that higher order multipoles are 
not a likely explanation, but we would like a more generic test of this
conclusion.  The basic problem is that most lenses provide only enough
data to constrain the standard ellipsoidal models, so more complex 
models that can explain the image fluxes are only constrained by
the degree to which the models are viewed as physical.

We noted in \S\ref{sec:parity} that there is a statistical tendency
for the anomalous  
flux ratios to appear as a suppression of the flux of the brightest saddle
points relative to the predictions of smooth models, and that this was
a characteristic property of low optical depth substructure.  It is
likely that any specific model which can explain the anomalies in 
individual lenses will be unable to reproduce this statistical property
for randomly selected lenses.  The worst case for examining this
expectation is the effect of the $m=4$ multipole, because it is the
largest amplitude higher order multipole with a symmetry matching a 
common image configuration, the cruciform quads like Q2237+0305.  
If an $m=4$ multipole is aligned with the ellipsoid it will 
preferentially affect the saddles as compared to the minima of
a cruciform quad, but it will symmetrically affect both saddles
and both minima.  If it is randomly misaligned with respect to the
ellipsoid, then it will not distinguish between saddles and minima,
and if the observed lens geometries are dominated by merging pairs
(fold catastrophes, like PG1115+080) or merging triples (cusp
catastrophes, like B1422+231), then it will again be unable to
distinguish saddles from minima.

We tested these hypotheses for models consisting of an ellipsoid 
with an aligned (boxy or disky) $a_4$ component or a randomly 
oriented $a_4$ component for a range of models for the lens 
population or typical lens geometry.  If we simulate $m=4$
models with the amplitudes typical of real elliptical galaxies or 
simulated CDM halos ($|a_4|\lesssim0.02$), the resulting 
$\sim 5\%$ flux perturbations are too small to be relevant for
explaining the anomalous flux ratios.  We must use models
of abnormally large amplitude, $|a_4|=0.05$, to produce
sufficiently large perturbations. 
After generating an image configuration, we fit the images using
our standard SIE $+$ shear models, fitting only the astrometric data
and ignoring the fluxes.  We then examined the statistics of
the flux residuals after dividing the images into magnification 
and parity subsamples just as in  \S\ref{sec:parity}.  
We focused on cusp and fold configurations defined by lenses 
with an opening angle between two images of $<30^\circ$.  Of the DK02
sample, the systems MG0414+0534, B0712+472, PG1115+080, B1422+231, and
B2045+265 would be classified as fold or cusp under this definition,
while B1608+656 and B1933+503 just fail to make the cut.

\begin{figure}  
\centerline{\psfig{figure=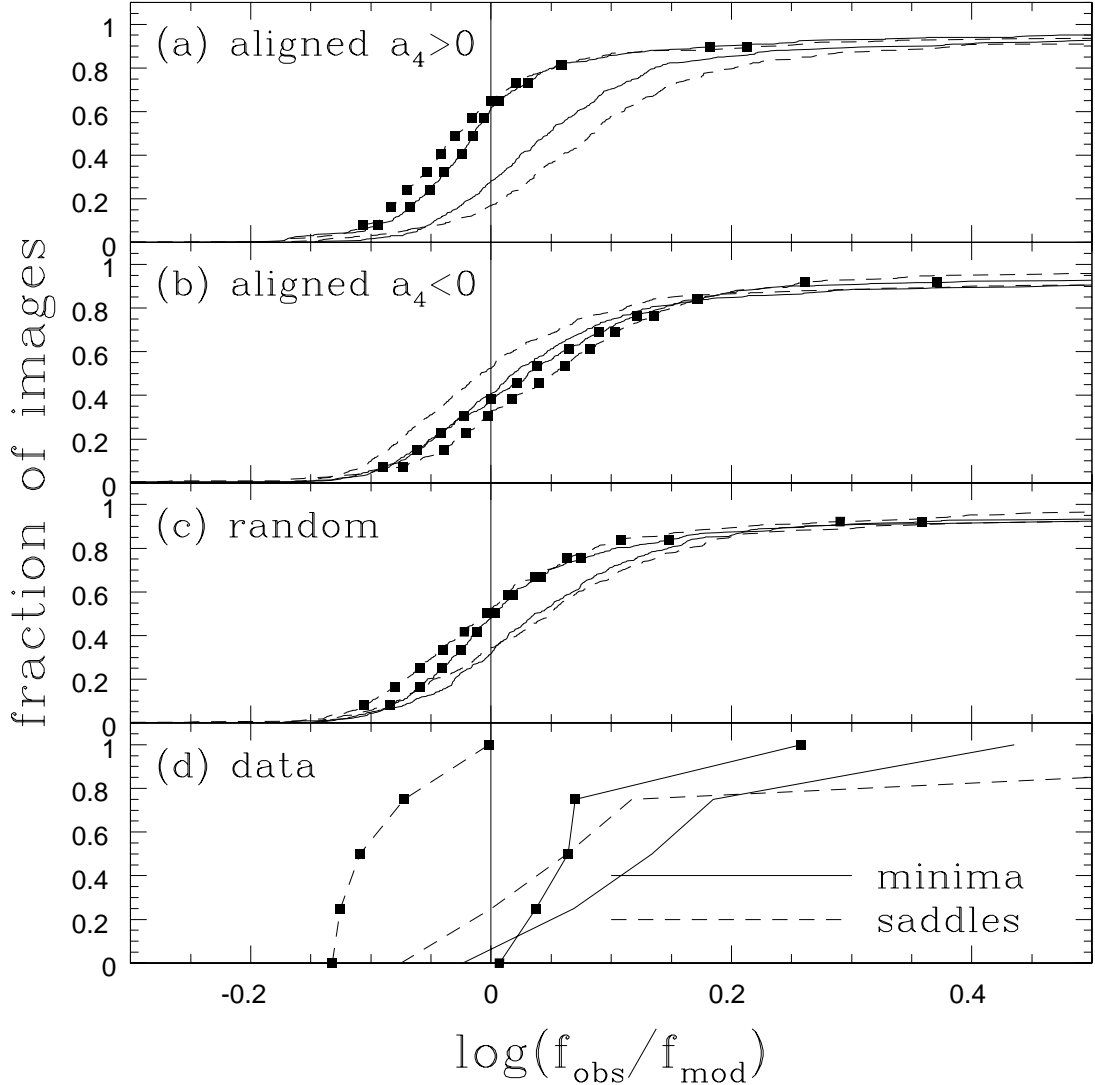,width=6.0in}}
\caption{Cumulative distributions of model flux residuals, 
$\log(f_{\rm obs}/f_{\rm mod})$, expected from Monte Carlo simulations
of fold and cusp lenses with $m=4$ components of amplitude
$|a_4|=0.05$.  Panels (a) and (b) correspond to perturbations aligned
with the major axis of the ellipsoid, for $a_4>0$ (disky) and $a_4<0$
(boxy) respectively.  Panel (c) shows results for randomly oriented
$m=4$ terms.  Panel (d) shows the flux anomalies observed for the
5 lenses in the DK02 sample satisfying the definition of fold/cusp 
described in the text.
The curve labeling is the same as in Figures \ref{fluxresid:monte} and
\ref{fluxresid:data}.
  \label{a4resid}
  }
\end{figure}

The results of our Monte Carlo calculation are displayed in Figure~\ref{a4resid}a-c.  
As expected, $a_4$ perturbations to fold or cusp
lenses do not systematically distinguish between image parities for
disky, boxy, or randomly oriented $m=4$ multipoles.  This is in stark
contrast to the properties of real fold and cusp lenses where the 
bright saddle points are preferentially demagnified compared to the
other images (Fig.~\ref{a4resid}d).  Clearly, $m=4$ perturbations are
incapable of producing this behavior, with KS test probabilities
of less than $0.002$ than any of these distributions agree with 
the distribution observed for the cusp/fold lenses in DK02.
We experimented with still higher order multipoles ($m>4$), and
found that they also fare poorly in both
systematically demagnifying saddles and differentiating between
saddles and minima.  The differences between distributions found
in the Monte Carlo simulations and in the real data mean that the
flux ratio anomalies cannot be due to higher order multipoles.  
{\it This would be true even if the models with higher order
multipoles could successfully model the anomalous flux ratios in
every lens!}  

\section{Ruling Out Microlensing of Radio Sources}

We are left only with substructure in the gravitational potential as a
viable explanation of the phenomenon.  Real lenses have substructures
on two mass scales, stars and satellite halos, so the last step of our
argument is to rule out stars as the source of the anomalous flux ratios
of lensed radio sources.  Microlensing is certainly an important 
phenomenon in the optical, where its effects are directly observed
as uncorrelated time variability in lensed quasars (e.g. Q2237+0305,
Wozniak et al.~\cite{Wozniak00}).  The principal 
arguments against significant microlensing in radio lenses are laid out 
by Koopmans \& de Bruyn (\cite{koopmans}). The key point is that the typical 
scale of the caustic network is set by the Einstein radii of the stars
($\theta_{\rm E}=3[M/M_\odot]^{1/2}\mu$as for a star of mass $M$, a lens
at $z_l=0.3$ and a source at $z_s=2$, hence the name microlensing), 
while the Compton limit for the minimum possible radio source size is
given by
\begin{equation}
  S(\nu) = \frac{2k T_b \nu^2}{(1+z)c^2}\pi \Delta\theta_{\rm min}^2.
\end{equation}
Table~\ref{compton} shows that the minimum source sizes for the 
four-image CLASS lenses are typically $\theta_{\rm min}\gtorder 10\mu$as,
or roughly 10 times larger than $\theta_{\rm E}$ for a stellar mass 
function dominated by low-mass stars.
A sufficiently non-thermal 
electron distribution can change these limits, but it is unlikely to reduce them
by an order of magnitude.   
While microlensing leads to fractional flux variations of order 
unity for sources smaller than the typical caustic size, larger sources
average over the magnification pattern and show far smaller variations.
For sources larger than $20 \mu$as we would expect $\lesssim 5\%$ 
fluctuations from stellar microlensing.  For microlensing to produce
the $\gtrsim 20\%$ flux anomalies of the radio lenses we would need
to have source sizes of order $\sim\mu$as.

One means of escaping the Compton limit on the radio source sizes is
to make most of the radio flux come from a relativistically beamed
source.  If a significant fraction of the source flux comes from a
source moving towards us at relativistic speeds, then the brightness
temperature of the beamed component can be increased by a
Doppler factor ${\cal D}=[\gamma (1-\beta\cos i)]^{-1}\sim\gamma$,
allowing a corresponding reduction in the source size.  For example,
there are $\sim 5\%$ uncorrelated changes in the radio fluxes of the
lensed images in B1600+434 that can be explained by microlensing
a beamed radio component with $\gamma \gtrsim 10$
(Koopmans \& de Bruyn \cite{koopmans}).  Much larger 
Lorentz factors would be needed for beaming to explain the 
$\gtrsim 20\%$ amplitude of the flux anomalies.  

In the absence of beaming, microlensing leads to changes in the image
flux ratios on time scales of years, while CDM substructure would 
lead to changes only on time scales of millenia or longer.  If there
is beaming, so that the characteristic velocities are superluminal
rather than simply the peculiar velocities of the lens, source and
observer, then the time scales will be $~1000$ times shorter.  Where
radio sources have been monitored for extended periods (e.g.
B1600+434, Koopmans \& de Bruyn \cite{koopmans}; 
B1933+503, Biggs et al.~\cite{biggs};
B1608+656, Fassnacht et al.~\cite{Fassnacht02}) the average
flux ratios show little time dependence.  While this is only weak
evidence against microlensing in the absence of beaming, it is
strong evidence against microlensing if there is beaming because
of the very short time scales created by the relativistic motions.  
Finally, if microlensing were important, we would expect a 
significant frequency dependence to the flux ratios because
we expect the source size to be a function of wavelength.  
As discussed in \S\ref{sec:opdepth}, this frequency
dependence has not been observed.

Because of the stress we have put on the dependence of the flux residuals
on image parity, we examined whether the differences between saddle points
and minima we discussed earlier persists for source sizes comparable to 
the Einstein radii of the substructure.  We ran inverse
ray-tracing simulations (e.g. Schneider et al.~\cite{Schneider92})
using 10240$^2$ rays and the particle-mesh (PM) force calculation
algorithm (Hockney and Eastwood~\cite{hockney}).  The PM method enjoys
several advantages for the microlensing problem, chiefly in the
simplicity and speed of the algorithm.  Using this code, we 
simulated microlensing at a saddle-point and at a minimum, each with an
unperturbed macro-model magnification of about 10. The saddle has a total
convergence and shear (coming from the macro model) of
$\kappa_0=0.525,\ \gamma_0=0.575$, while the minimum has
$\kappa_0=0.475,\ \gamma_0=0.425$.  We lay down a random distribution
of microlenses with Einstein radii $\theta_{\rm E}=1\mu$as and a number density
corresponding to convergence $\kappa_\ast=0.075$.  The remaining surface
density, $\kappa_0-\kappa_\ast$, was in a smooth component.

\begin{figure}
\centerline{\psfig{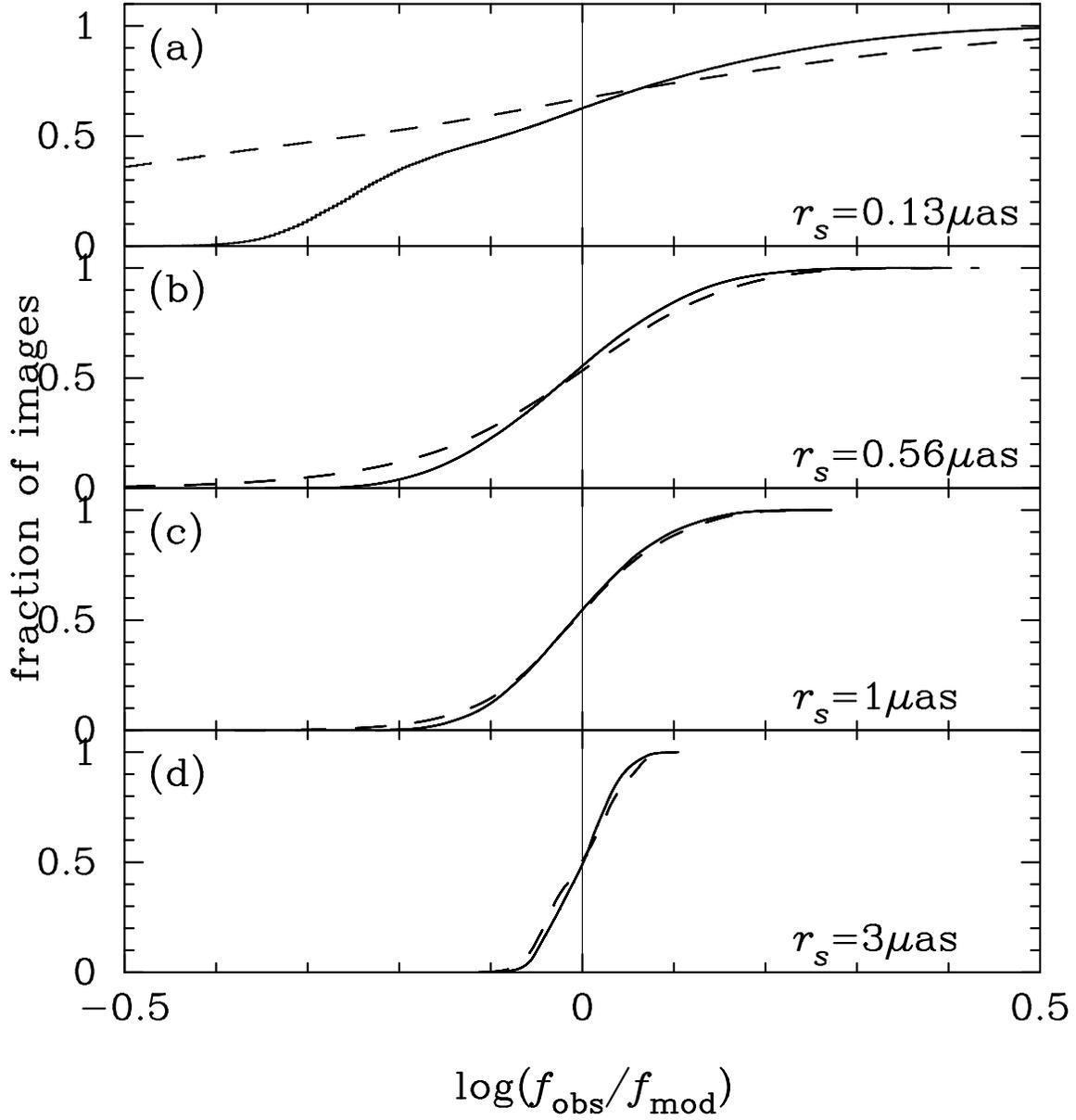}}
\caption{Cumulative distributions of magnification perturbations for microlensing,
as a function of source size $r_s$.  The solid lines correspond to a
minimum with $\kappa_0=0.475$ and $\gamma_0=0.425$ while the dashed
lines correspond to a saddle point with $\kappa_0=0.525$ and
$\gamma_0=0.575$.  The simulations used microlenses with Einstein
radii of $\theta_{\rm E}=1\mu$as and surface density
$\kappa_\ast=0.075$. }
\label{microlensing}
\end{figure}

In Figure~\ref{microlensing} we plot the distribution of flux residuals
as a function of source size. For a source small compared to $\theta_{\rm E}$,
like the case $r_s=0.13\mu$as corresponding to our grid resolution, we
see that the magnification distribution for the saddle point is skewed
towards demagnification compared to that for the minimum, in agreement
with previous results (e.g. Schechter \& Wambsganss~\cite{sw02}; see their Fig.~3).
For larger sources, which we illustrate for the cases $r_s=\theta_{\rm E}=1\mu$as
and $3\theta_{\rm E}=3\mu$as, the widths of the distribution narrow rapidly
(with $\sigma_\mu/\mu_0\propto 1/r_s$ roughly, e.g. Koopmans \& de Bruyn \cite{koopmans}) 
and the differences between the 
distributions for saddle points and minima vanish.  The mean 
$\langle\log_{10}\mu/\mu_0\rangle$ provides a measure of the
skewness, and for source sizes of $r_s=0.13$, $0.56$, $1$ and $3\mu$as we find
values of $\langle\log_{10}\mu/\mu_0\rangle=-0.26$, $-0.025$
$-0.01$ and $-0.002$ respectively.
Evidently, even source sizes as small as $\theta_{\rm E}/2=0.5\mu$as are sufficient to
erase the asymmetry in the distribution of perturbations to
saddle-point images.  In reality, radio QSO's are expected to have
much larger angular sizes (10--30$\mu$as, see Table~\ref{compton}), 
giving even narrower, more symmetric
distributions than those plotted in Figure~\ref{microlensing}.

\section{Summary and Discussion}

We have considered three alternatives to substructure as the source of
the flux anomalies observed in four-image radio lenses: propagation effects
like refractive scattering in the interstellar medium (ISM) of the
lens galaxy, problems in the macro model describing the gravitational
potential of the lens galaxy, and microlensing by ordinary stars in
the lens galaxy.  We find that none of these alternatives are likely
to explain the phenomenon.

Generating the flux anomalies using the ISM fails on two counts.  First,
the anomalies in the flux ratios of the radio lenses are essentially
independent of wavelength, unlike any normal propagation effect in
the ISM (scintillation, refractive scattering, free-free absorption).
Any attempt to use the ISM as an explanation of the anomalous flux ratios 
must find a mechanism to produce a nearly frequency independent optical 
depth -- the radio equivalent of the ``gray dust'' proposals for Type Ia
supernovae (Aguirre~\cite{Aguirre98}).  Second, the flux ratio anomalies
depend on the image parity and magnification. In particular, the brightest
saddle point image has a tendency to be demagnified relative to the
other images, which is a characteristic of low optical depth substructure
(Schechter \& Wambsganss~\cite{sw02}, Keeton~\cite{Keeton02}).  Since 
the ISM should preferentially perturb the least magnified images, which will
be the most compact, and show no dependence on the image parity, we
can statistically rule out any propagation effect. 

Solving the flux anomalies by modifying the smooth gravitational potential
of the lens also fails on two counts.  Following other studies (Evans \&
Witt~\cite{Evans02}, M\"oeller, Hewett \& Blain~\cite{Moeller02},
Quadri, M\"oeller \& Natarajan~\cite{Quadri02}), we investigated adding
higher order multipoles ($\phi \propto \cos m\theta$  
with $m > 2$) to the potential. For the models
we attempted (which were restricted to adding $m=3$ and $m=4$) 
we could find solutions without flux anomalies for only one of the six
lenses from DK02 
which had anomalies in our standard ellipsoidal lens models.
The fits required the higher order multipoles to be significantly misaligned
with respect to the axes of the ellipsoid and had perturbation amplitudes
larger than seen in real galaxies or halo simulations.  In three of
the lenses, the presence of additional constraints, either Einstein
ring images of the quasar host  
galaxy or additional lensed components of the radio source, allowed us
to measure directly the magnitude of the higher multipoles in the
gravitational potential.  In every instance, the $m=3$ and $4$ multipoles
were constrained to the small amplitudes typically observed in galaxies
and halo models, largely justifying the use of ellipsoidal mass models to describe
lens galaxies.  

The second problem with the higher order multipole models is that they 
generally cannot reproduce the observed parity dependence of the flux 
anomalies.  They can do so in very specific cases where the symmetry
of the lens (a cruciform quad like Q2237+0305) matches the symmetry of
the 
multipole (a $\cos 4\theta$ multipole), and the multipole has very
large amplitude and is aligned with 
the dominant ellipsoidal potential.  However, none of the radio quads
with anomalies have this symmetry (they generally have merging image
pairs or triples associated with fold and cusp caustics) and the 
models  require multipoles that are not
aligned with the ellipsoid.  If we examine the distribution of anomalies
found by fitting lenses with higher order multipoles using only standard
ellipsoidal lens models, we find that the statistical properties of the
brightest saddle point do not stand out from those of the other images.
This allows us to rule out these models as an explanation even if they
could fit the anomalies in each individual lens (which they cannot).

The failure of the ISM or changes in the smooth potential to explain the
data means that the explanation must be substructure in the potential.
These substructures must either be dark objects,
since luminous satellites are not abundant enough to
account for the preponderance of anomalous fluxes (e.g. Chiba~\cite{Chiba02}), 
or the stellar populations in the lens galaxy.  Either source of 
substructure will lead to the statistical differences between saddle
points and minima if the optical depth is low 
(Schechter \& Wambsganss~\cite{sw02}, 
Keeton~\cite{Keeton02}).  The angular scales of the magnification 
patterns are very different for stars (microarcseconds) and satellites
(milliarcseconds) and by using radio sources we should be averaging
out the stellar magnification patterns to leave only the contribution
from the satellites.  We tested this by taking typical microlensing
patterns for saddle point and minimum images, which show the expected
differences for point sources, and smoothing on larger and larger
scales.  As expected, the magnification distributions for the saddle 
point and the minimum show significant differences only when the source 
size is smaller than the Einstein radius of the stars.  Source sizes
of even one microarcsecond are sufficient to eliminate the differences,
so radio sources could have angular sizes even an order of magnitude
smaller than the Compton limit without microlensing making a significant
contribution to the flux anomalies of radio lenses.

In summary, CDM substructure remains the
best explanation of the flux ratio anomalies of gravitational lenses. 
The most powerful piece of evidence is the statistically significant
difference between the anomalies in saddle points and minima and
between the bright and faint images.  Substructure makes a very 
specific prediction that the brightest saddle point should show a
very different distribution from the other images, as observed in the
data, that is very difficult or impossible to reproduce using either
the ISM or changes in the smooth potential.

However, the issues raised by these considerations point to future
observations that can further clarify the origin of anomalous fluxes
in lens systems.  ISM effects can be more strongly constrained by 
measuring flux ratios at still higher frequencies (e.g. 43~GHz at the
VLA).  Mid-infrared (5--10$\mu$m) flux 
ratios, where the wavelength is far too short to be bothered by electrons and 
far too long to be bothered by dust, are difficult to measure but completely
insensitive to the ISM (e.g. Agol, Jones \& Blaes~\cite{Agol00}).
 The mid-infrared source sizes should also be 
large enough to be affected only by satellites, adding a further check
for whether we can be misinterpreting microlensing effects as substructure. 
Integral field spectroscopy, to compare the flux ratios in the optical
continuum to those in the broad emission lines, provides another way to
separate the effects of microlensing and substructure because the broad
line emitting regions should be significantly larger than the source of
the optical continuum (e.g. Moustakas \& Metcalf~\cite{Moustakas02}).
Monitoring the lenses, either in the radio or the optical, to look for
changes in the flux ratios, can also be used to distinguish microlensing
effects from CDM substructure.  
Observations to find additional lensed structures 
are the best approach to determining whether
more complicated lens potentials are needed.  Clean constraints on more
complicated angular structures can be obtained by analyzing the shapes
of the Einstein ring images of host galaxies found in deep, high-resolution
infrared images (see Kochanek et al.~\cite{Kochanek01}).

\acknowledgments{We thank Roger Blandford, Scott Gaudi, Gil Holder,
Chuck Keeton, Feryal \"Ozel and David Rusin for helpful discussions.  
We also thank Chuck Keeton
for rapidly adding the higher order multipole potentials to the
{\it lensmodel} package, and Chris Fassnacht for giving us the
unpublished 5 and 15~GHz flux measurements for B1608+656. ND 
acknowledges the support of NASA through Hubble Fellowship grant
\#HST-HF-01148.01-A awarded by STScI, which is operated by AURA
for NASA, under contract NAS 5-26555.
CSK is supported by the Smithsonian Institution and NASA grant
NAG5-9265.}


\begin{thebibliography}{}

\bibitem[2000]{Agol00}
  Agol, E., Jones, B., \& Blaes, O., 2000, ApJ, 545, 657

\bibitem[1998]{Aguirre98} 
  Aguirre, A.\ N.\ 1998, ApJ, 512, L19

\bibitem[1997]{Bade97}
  Bade, N., Siebert, J., Lopez, S., Voges, W., \& Reimers, D., 1997,
    A\&A, 317, 13

\bibitem[1989]{bender}
  Bender, R.\ et al.\ 1989, A\&A, 217, 35

\bibitem[2000]{biggs}
Biggs, A.\ D.\ et al.\ 2000, MNRAS 318, 73

\bibitem[2001]{Bode01}
  Bode, P., Ostriker, J. P. and Turok, N. 2001, ApJ, 556, 93

\bibitem[2002]{Bradac02}
  Bradac, M., Schneider, P., Steinmetz, M., Lombardi, M., \& King, L.J., 2002,
   A\&A, 388, 373

\bibitem[2003]{Burkert03}
  Burkert, A., \& Naab, T., 2003, in Galaxies \& Chaos, eds. G.
    Contopulos \& N. Voglis (Springer Verlag) [astro-ph/0301385]

\bibitem[2002]{chen}
Chen, J., Keeton, C. R.\ \& Kravtsov, A.\ 2002, in preparation

\bibitem[2002]{Chiba02}
Chiba, M.\ 2002, ApJ, 565, 17

\bibitem[2001]{Cohn01}
  Cohn, J.D., Kochanek, C.S., McLeod, B.A., \& Keeton, C.R., 2001, ApJ,
    554, 1216

\bibitem[2002]{dk02}
Dalal, N.\ \& Kochanek, C.\ S.\ 2002, ApJ, 572, 25

\bibitem[2001]{ew}
Evans, N.\ W.\ \& Witt, H.\ J.\ 2001, MNRAS, 327, 1260

\bibitem[2002]{Evans02}
Evans, N.\ W.\ \& Witt, H.\ J.\ 2002, astro-ph/0212013

\bibitem[1998]{Falco99}
  Falco, E.E., Impey, C.D., Kochanek, C.S., Lehar, J., McLeod, B.A.,
   Rix, H.-W., Keeton, C.R., Munoz, J.A., \& Peng, C.Y., 1999,
   ApJ, 523, 617

\bibitem[1996]{Fassnacht96}
   Fassnacht, C.D., Womble, D.S., Neugebauer, G., Browne, I.W.A., Readhead,
   A.C.S., Matthews, K., \& Pearson, T.J., 1996, ApJL, 460, 103

\bibitem[1999]{Fassnacht99}  
  Fassnacht, R.D., Blandford, R.D., Cohen, J.G., et al., 1999, AJ, 117, 658 

\bibitem[2002]{Fassnacht02}
  Fassnacht, R.D., Xanthopoulos, E., 
   Koopmans, L.V.E., \& Rusin, D., 2002, ApJ, 581, 823

\bibitem[1992]{Hewitt92}
  Hewitt, J.N., Turner, E.L., Lawrence, C.R., Schneider, D.P., \& Brody,
  J.P., 1992, AJ, 104, 968

\bibitem[1994]{heyl}
Heyl, J.\ S., Hernquist, L.\ \& Spergel, D.\ N.\ 1994, ApJ, 427, 165

\bibitem[1981]{hockney}
Hockney, R. W. and Eastwood, J. W. 1981, {\it Computer Simulation
Using Particles}, (McGraw-Hill: New York)

\bibitem[1985]{Huchra85}
  Huchra, J., Gorenstein, M., Kent, S., Shapiro, I., Smith, G.,
   Horine, E., \& Perley, R., 1985, AJ, 90, 691

\bibitem[1998]{impey98}
  Impey, C.D., Falco, E.E., Kochanek, C.S., Lehar, J., McLeod, B.A.,
    Rix, H.-W., Peng, C.Y., \& Keeton, C.R., 1998, ApJ, 509, 551

\bibitem[1998]{Jackson98}
  Jackson, N., Nair, S., Browne, I.W.A., Wilkinson, P.N., Muxlow, T.W.B.,
  de Bruyn, A.G., Koopmans, L.V.E., et al., 1998, MNRAS, 296, 483

\bibitem[1997]{Katz97}
  Katz, C.A., Moore, C.B., \& Hewitt, J.N., 1997, ApJ, 475, 512

\bibitem[1993]{Kauffmann93}
Kauffmann, G., White, S.D.M., \& Guiderdoni, B., 1993, MNRAS, 264, 201

\bibitem[2001]{Keeton01}  
Keeton, C.\ R.\ 2001a, astro-ph/0111595

\bibitem[2001]{Keeton01b}  
Keeton, C.\ R.\ 2001b, astro-ph/0102341

\bibitem[2002]{Keeton02}  
Keeton, C.\ R.\ 2002, astro-ph/0209040

\bibitem[2002]{Keetonetal02}
Keeton, C.R., Gaudi, B.S., \& Petters, A.O., 2002, ApJ submitted, astro-ph/0210318 

\bibitem[1969]{kellermann}
Kellermann, K.\ I.\ \& Pauliny-Toth, I.\ I.\ K.\ 1969, ApJ, 155, L71

\bibitem[1999]{Klypin99}
  Klypin, A., Kravtsov, A.V., Valenzuela, O., \& Prada, F., 1999, ApJ, 522, 82.

\bibitem[1991]{Kochanek91}
  Kochanek, C.S., 1991, ApJ, 373, 354

\bibitem[2001]{Kochanek01} 
  Kochanek, C.~S., Keeton, C.~R., \& McLeod, B.~A.\ 2001, \apj, 547, 50 

\bibitem[2002]{Kochanek01b}Kochanek, C.S., 2002, in the proceedings of the Yale 
  Cosmology Workshop, P. Natarajan, ed.  [astro-ph/0106495]

\bibitem[2000]{koopmans}
  Koopmans, L.\ V.\ E.\ and de Bruyn, A.\ G.\ 2000, A\&A, 358, 793

\bibitem[1988]{Magain88}
  Magain, P., Surdej, J., Swings, J.-P., Borgeest, U., \& Kayser, R.,
    1988, Nature, 334, 325

\bibitem[1998]{Mao98}
  Mao, S., \& Schneider, P., 1998, MNRAS, 295, 587

\bibitem[1999a]{Marlow99a}
  Marlow, D.R., Browne, I.W.A., Jackson, N., \& Wilkinson, P.N., 
    1999a, MNRAS, 305, 15

\bibitem[1999b]{Marlow99b}
  Marlow, D.R., Myers, S.T., Rusin, D., Jackson, N., Browne, I.W.A., Wilkinson, P.N.,
  Muxlow, T., Fassnacht, C.D., Lubin, L., Kundic, T., Blandford, R.D., Pearson, T.J.,
  Readhead, A.C.S., Koopmans, L., \& de Bruyn, A.G., 1999b, AJ, 118, 654

\bibitem[2001]{Metcalf01}
  Metcalf, R.B., \& Madau, P., 2001, ApJ, 563, 9

\bibitem[2002]{mz02}
Metcalf, R. B. and Zhao, H. 2002, ApJ, 567, L5

\bibitem[2002]{Metcalf02}
Metcalf, R. B. 2002, astro-ph/0203012

\bibitem[1967]{Mezger67}
  Mezger, P.C., \& Henderson, A.P., 1967, ApJ, 147, 471

\bibitem[2002]{Moeller02}
  M\"oeller, O., Hewett, P., Blain, A.\ W.\ 2002, astro-ph/0212467

\bibitem[1999]{Moore99}
  Moore, B. Ghigna, S., Governato, F., Lake, G.,
   Quinn, T., Stadel, J., \& Tozzi, P., 1999, ApJ, 524, L19

\bibitem[2002]{Moustakas02}
  Moustakas, L.A., \& Metcalf, R.B., 2002, astro-ph/0206176

\bibitem[2001]{munoz01}
   Munoz, J.A., Kochanek, C.S., \& Keeton, C.R., 2001, ApJ, 558, 657

\bibitem[1999]{Myers99}
  Myers, S.T., Rusin, D., Fassnacht, C.D., Blandford, R.D., Pearson, T.J.,
  Readhead, A.C.S., Jackson, N., Browne, I.W.A., Marlow, D.R., Wilkinson, P.N.,
  Koopmans, L.V.E., \& de Bruyn, A.G., 1999, AJ, 117, 2565

\bibitem[1992]{narayan92}
  Narayan, R.\ 1992, Phil.\ Trans.\ Roy.\ Soc., 341, 151

\bibitem[1992]{Patnaik92}
  Patnaik, A.R., Browne, I.W.A., Walsh, D., Chaffee, R.H., \& Foltz, C.B., 1992,
   MNRAS, 254, 655P

\bibitem[1999]{Patnaik99} 
  Patnaik, A.R., Kemball, A.J., Porcas, R.W., \& 
   Garrett, M.A., 1999, MNRAS, 307, L1

\bibitem[2001]{Patnaik01} 
  Patnaik, A.R. \& Narasimha, D., 2001, MNRAS, 326, 1403

\bibitem[2000]{Phillips00}
  Phillips, P.M., Norbury, M.A., Koopmans, L.V.E., Browne, I.W.A., Jackson, N.J.,
  Wilkinson, P.N., Biggs, A.D., Blandford, R.D., de Bruyn, A.G., Fassnacht, C.D.,
  et al., 2000, MNRAS, 319, 7P

\bibitem[2002]{Quadri02}
  Quadri, R., M\"oeller, O., \& Natarajan, P., 2002, ApJ submitted, astro-ph/0212467

\bibitem[2002]{Reimers02} 
  Reimers, D., Hagen, H.-J., Baade, R., Lopez, S., \& Tytler, D.,
   2002, A\&A, 382, 26

\bibitem[2001]{Rest01}
Rest, A., van den Bosch, F.C., Jaffe, W., Tran, H., 
  Tsvetanov, Z., Ford, H.C., Davies, J., \& Schafer, J.L, 2001, AJ, 121, 2431

\bibitem[2000]{Ros00}
  Ros, E., Guirado, J.C., Marcaide, J.M., Perez-Torres, M.A., Falco, E.E., Munoz,
  J.A., Alberdi, A., \& Lara L., 2000, A\&A, 362, 845

\bibitem[2002]{sw02}
Schechter, P.\ L. \& Wambsganss, J.\ 2002, \apj, 580, 685

\bibitem[1992]{Schneider92}
  Schneider, P., Ehlers, J., \& Falco, E.E., 1992, Gravitational Lenses,
   (Springer Verlag: Berlin)

\bibitem[1998]{Sykes98} 
  Sykes, C.M., Browne, I.W.A., Jackson, N.J., Marlow, D.R., Nair, S., 
  Wilkinson, P.N., Blandford, R.D., Cohen, J., Fassnacht, C.D.,
  Hogg, D., Pearson, T.J., Readhead, A.C.S., Womble, D.S., Myers, S.T.,
  de Bruyn, A.G., Bremer, M., Miley, G.K., \& 
  Schilizzi, R.T., 1998, MNRAS, 301, 310

\bibitem[2000]{Trotter00}
  Trotter, C.S., Winn, J.N. \& Hewitt, J.N., 2000, ApJ, 535, 671

\bibitem[1998]{walker98}
Walker, M.\ A.\ 1998, MNRAS, 294, 307

\bibitem[1980]{Weymann80}
   Weymann, R.J., Latham, D., Roger, J., Angel, P.,
   Green, R.F., Liebert, J., Turnshek, D.A., Turnshek, D.E.,
   \& Tyson, J.A., 1980, Nature, 285, 641

\bibitem[1999]{Wisotzki99}
   Wisotzki, L., Christlieb, N., Liu, M.C., Maza, J., Morgan, N.D.,
    \& Schechter, P.L.,  1999, A\&A, 348, 41

\bibitem[2002]{Wisotzki02}
   Wisotzki, L., Schechter, P.L., Bradt, H.V., Heinmuller, J.,
     Reimers, D., 2002, A\&A, 395, 17
    \& Schechter, P.L.,  1999, A\&A, 348, 41


\bibitem[2000]{Wozniak00} 
   Wozniak, P.R., Udalski, A., Szymanski, M., et al.,
    2000, ApJL, 540, 65

\bibitem[2002]{Zentner02}
  Zentner, A.R., \& Bullock, J.S., 2002, astro-ph/0212339

\end{thebibliography}
\end{document}